\documentclass[draft,sw]{AGUTeX}

 \usepackage{lineno}
 \usepackage{footnote}

  \usepackage[xdvi]{graphicx}
  \usepackage{url}
  \setkeys{Gin}{draft=false}
\authorrunninghead{GUERRA ET AL.}

\titlerunninghead{ENSEMBLE FORECASTING}

\authoraddr{Corresponding author: J. A. Guerra,
The Catholic University of America,
200 Hannan Hall, Washington D.C.,20064, USA.
(94guerraagui@cardinalmail.cua.edu)}

\begin{document}

%
%

\title{Ensemble Forecasting of Major Solar Flares -- First Results}
%
%

%
%



\authors{J. A. Guerra,\altaffilmark{1,2},  A. Pulkkinen \altaffilmark{3}, and V. M. Uritsky \altaffilmark{1,2}}

\altaffiltext{1}{Physics Department, The Catholic University of America, Washington DC, USA.}

\altaffiltext{2}{Heliophysics Science Division, NASA GSFC, Greenbelt, MD, USA.}

\altaffiltext{3}{Space Weather Laboratory, Heliophysics Science Division, NASA GSFC, Greenbelt, MD, USA.}




%
%


\begin{abstract}
We present the results from the first ensemble prediction model for major solar flares (M and X classes). The primary aim of this investigation is to explore the construction of an ensemble for an initial prototyping of this new concept. Using the probabilistic forecasts from three models hosted at the Community Coordinated Modeling Center (NASA-GSFC) and the NOAA forecasts, we developed an ensemble forecast by linearly combining the flaring probabilities from all four methods. Performance-based combination weights were calculated using a Monte-Carlo-type algorithm that applies a decision threshold $P_{th}$ to the combined probabilities and maximizing the Heidke Skill Score (HSS). Using the data for 13 recent solar active regions between years 2012 - 2014, we found that linear combination methods can improve the overall probabilistic prediction and improve the categorical prediction for certain values of decision thresholds. Combination weights vary with the applied threshold and none of the tested individual forecasting models seem to provide more accurate predictions than the others for all values of $P_{th}$. According to the maximum values of HSS, a performance-based weights calculated by averaging over the sample, performed similarly to a equally weighted model. The values $P_{th}$ for which the ensemble forecast performs the best are 25 \% for M-class flares and 15 \% for X-class flares. When the human-adjusted probabilities from NOAA are excluded from the ensemble, the ensemble performance in terms of the Heidke score, is reduced.
\end{abstract}

%
%

%

\begin{article}

%
%

\section{Introduction}

Forecasting solar flares is perhaps one of the greatest challenges in the Heliophysical sciences. Modelers face the problem of utilizing quantities and parameters usually measured from the instantaneous active region photospheric magnetic field -- which contains limited information on the active region's flare production \citep{BarnesLeka2008}. Moreover,  predictions from any model are often subjected to biases due to the method's training process and the statistical sample that was employed. When examining forecasts from different models or methods, it is usual to find that for the same condition of the photospheric magnetic field, they can give varying values for probabilities of a particular flare to happen. This variability becomes a reality on a daily basis for Space Weather forecasters. In the decision making process, forecasters can use as many pieces of information as they need to ensure that their choices will translate into reducing risk and costs on those system vulnerable to the solar activity.

The combination of forecasts is an approach that has been widely used in almost every discipline in which forecasts are important (see \cite{Armstrong2001} for an extensive review of literature on this topic). It has been proven that combining forecasts improves the accuracy by reducing the uncertainties associated with data imperfections, biases, or model approximations \citep{Armstrong2001,Clemen1989,Genre2013}. Generally, an improved forecast can be achieved if the combination components contain useful and independent information about the system they forecast. On the other hand, improvement greatly depends on numerous factors or internal parameters of the combination method: number of forecasts to be combined, nature of the forecast (expert assessment, extrapolations), and the parameter that quantifies the difference between forecast and observations, to name a few.

Using forecast combinations is known as {\it ensemble forecasting}, and there are different ways it can be performed. One type of ensemble often employed in climate forecasting involves generating several forecasts using the same method by perturbing the initial conditions within the uncertainties of the observational data. In this way, the uncertainty of the prediction can be assessed and reduced \citep{Collins2007}. In climatology, it is also common to use ensembles that encompass the combination of predictions from different times in order to make more accurate forecasts \citep{Armstrong2001}. Another ensemble can be constructed using forecasts given by different methods. This report will concentrate in this type of ensemble forecasts.

When combining forecasts made using different methods or techniques, it is strongly recommended to use simple combination schemes \citep{Armstrong2001,Clemen1989}. In particular, using linear combinations of forecasts with equal and unequal combination weights has proven to be straightforward and successful in certain research disciplines, {\it e.g.} economics. \cite{Armstrong2001} suggested the use of unequal weights when there is enough evidence to support it, for instance, when forecasts are statistically biased  \citep{GrangerandRamanathan1984}. He also proposed that weights can be calculated from track records (showing how well each method has performed in forecasting previous events) as the inverse value of some error measure.

In this article, we explore for the first time the idea of making a prediction for major solar flares (M and X classes) using an ensemble of methods. The aim of this work is to propose a specific method for constructing an ensemble forecast and to find the conditions (method internal parameters) for which the ensemble performs better than any of its members. In the following sections we will describe the methods employed in this study (Section 2) and how the method predictions are combined (Section 3). Section 4 describes the performance-based constructions of the ensemble. Description of the Monte Carlo type algorithm utilized in the constructions of the ensemble model is provided in Section 5. In Section 6 we discuss the main results of our work. Section 7 summarizes our findings and remarks on the future work. Appendix A contains the description of the active regions used in the analysis and provides additional details on time series generated by the methods.

\section{Solar Flares Forecast Methods}
In this article we will use four flare forecasting methods for constructing the ensemble forecast. The models are: the Magnetic Forecast (MAG4; U. of Alabama - Huntsville), the Automatic Solar Synoptic Analyzer (ASSA; Korean Space Weather Center), the Automated Solar Activity Prediction (ASAP; U. of Bradford - UK), and the forecasts given by the NOAA Space Weather Prediction Center (NOAA). The Community Coordinated Modeling Center (CCMC, \url{ccmc.gsfc.nasa.gov}) at NASA's Goddard Space Flight Center hosts the first three of these models, which are fully automated. All four models are based on the same basic idea: the instantaneous spatial configuration of the active region photospheric magnetic field provides some information about the occurrence of solar flares in the future which can be used for flare prediction. The models MAG4, ASSA, and ASAP report probabilistic forecasts in near real-time (15 - 60 min cadence) while NOAA reports every 24 hours. They process full-disk photospheric imagery (magnetograms and continuum), identifying all relevant ARs and strong field regions, and subsequently calculating probabilities for flaring of each AR and the full disk. For all these models, flaring probabilities are given for a prediction window of 12 or 24 hours from the forecast time. Each model has been trained using a different technique and algorithms, and therefore it is expected that for the same photospheric conditions their forecasts are different (in some cases quiet significantly, see Figure \ref{time_series}) from each other. It is the discrepancy in the forecasting of the same events that motivated us to construct an ensemble prediction capable of using the advantages the individual methods and combine them into a more accurate forecasting system. Below is a more detailed description of the four methods included into our ensemble forecast.

 \begin{figure}
 \centering
 \noindent\includegraphics[width=30pc]{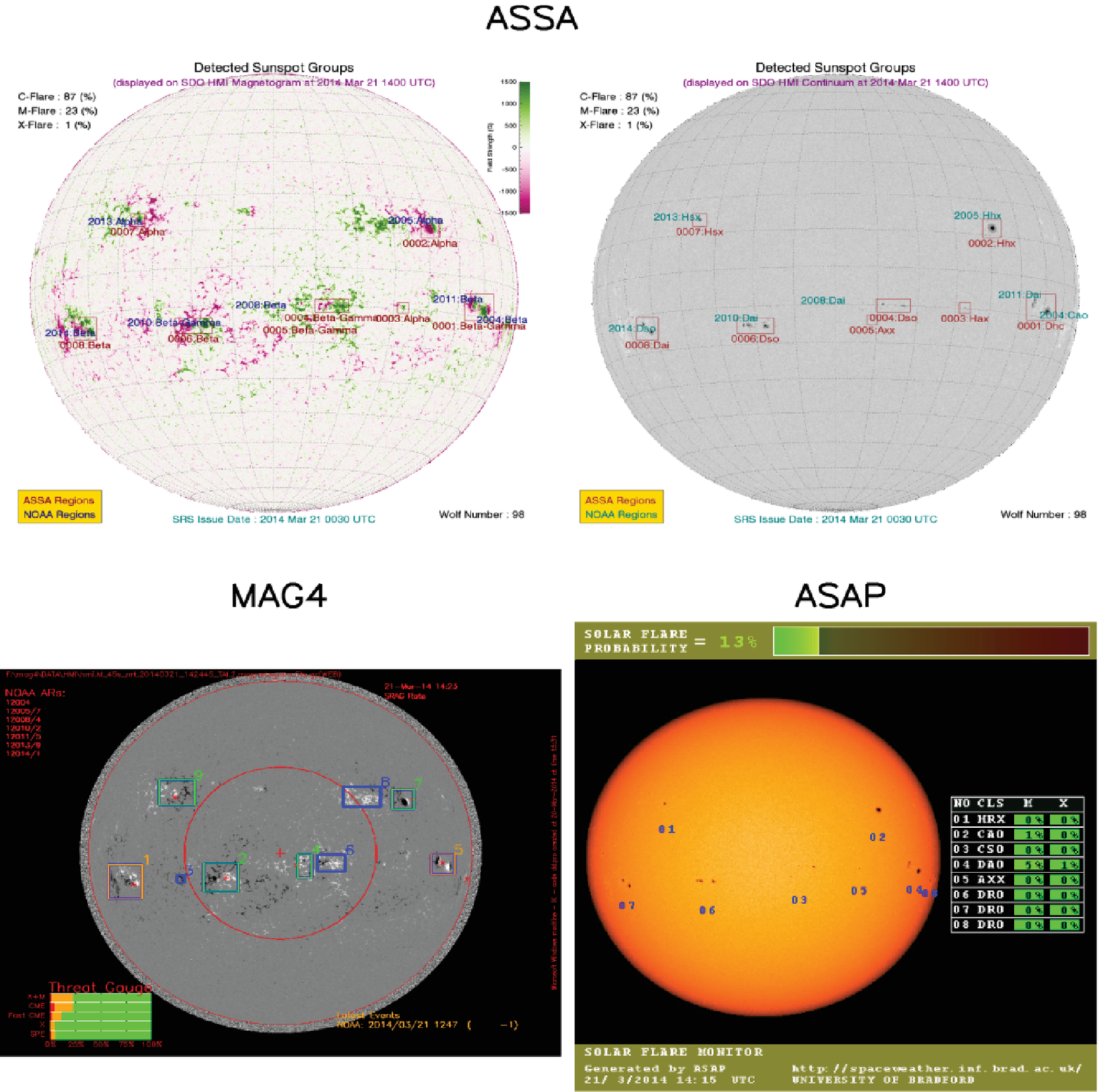}
 \caption{{\small Example of graphical output from the flare forecasting methods hosted at the Community Coordinated Modeling Center (CCMC). Top panels: ASSA forecast displaying all identified active regions on HMI magnetograms (left) and continuum (right). Full-disk flaring probabilities (upper left corner) are calculated from active region's flaring probabilities assuming Poisson distribution of events. Bottom-left panel: MAG4 forecast showing detected active regions (NOAA and other non-NOAA). MAG4 provides a threat gauge (calculated from active region predictions and their uncertainties) in which green bars represents a ``all-clear" conditions for solar events. Bottom-right: ASAP forecast providing flaring probabilities for M and X class flares for individual active regions and the full disk.}}
 \label{forecasting_methods}
 \end{figure}

\subsection{MAG4}

MAG4 was developed at the University of Alabama - Huntsville with support from the Space Radiation Analysis Group at Johnson Space Flight Center (NASA/SRAG) for forecasting M and X class flares, CMEs, fast CME, and Solar Energetic Particle events (\cite{Falconer2014}). This method uses the magnitude of the transverse gradient ($^{L}$WL$_{SG}$) of the line-of-sight magnetic field integrated over all of polarity inversion lines present in strong field areas ($>$ 150 G) as a proxy for the active region free magnetic energy \citep{Falconer2003,Falconer2014}. In particular, using forecasting curves - an empirical relation between values of $^{L}$WL$_{SG}$ and the flare class production rate ($R$) for a 24-hour forward window - MAG4 predicts the combined number of major flares (M \& X) and the number of  X-class flares alone. Event rates (R) are transformed into probability $P$ values by assuming a Poisson statistics yielding $P=1-\exp(-{\rm R}\Delta t)$, where $\Delta t$ is the prediction window. MAG4 reports forecasts at a 60-min cadence. 

Forecasts from MAG4 are affected by the projection effects present in active regions located beyond 30 heliocentric degree. MAG4 developers are currently updating the software for using vector magnetograms, which will improve forecasts outside the 30-heliocentric-degree circle. The current version of MAG4 also provides a second forecast which considers the active region flaring history (free-energy + flares; see \cite{Falconer2014} for more details). For active regions that have previously produced flares, the forecast curves (event rate vs $^{L}$WL$_{SG}$) are different, and therefore, the predicted event rates. In this investigation we will use the MAG4 forecasts based on the free-energy proxy parameter only.

\subsection{ASSA}

The ASSA code consists of 3 modules: (1) sunspot group identification and classification, (2) coronal hole detection, and (3) filament detection. It was developed at the Korean Space Weather Center, part of the National Radio Research Agency, Republic of Korea. The first module processes SDO/HMI images (magnetograms and continuum) and performs the McIntosh and Mt. Wilson classifications of all detected active regions. Flare probabilities for each active region are calculated using Poisson statistics based on the average flare rates for its McIntosh class. Average flare rates were determined by analyzing historical data of AR McIntosh classes and GOES X-ray data from 1996 to 2011 (see table 1 in \citep{assa_manual} page 13). Full disk probability is calculated from the active region probabilities as $P_{\rm fd}=1-(1-P_{1})(1-P_{2})...(1-P_{N})$, where $N$ is the number of active regions on the solar disk. All probabilities given by ASSA correspond to a forecasting window of 12 h. Refer to \url{http://www.spaceweather.go.kr/assa/} for further details on this model.

\subsection{ASAP}

This forecasting model was developed at the University of Bradfrod, UK. It was originally trained with SOHO/MDI continuum and LOS magnetograms \citep{ColakQahwaji2008} but has been recently updated to process the same data from SDO/HMI. Similarly to ASSA, ASAP identifies sunspot groups and classifies them according to the McIntosh class. The area of each sunspot group and its McIntosh class are used as inputs to an algorithm in order to generate probabilistic predictions for flares in the next 24 hours. This algorithm uses two neural-network systems that were trained using a catalog of solar events with more than 70,000 flares from 1982 to 2006 \citep{ColakQahwaji2009}. The first system computes the probability of having a flare of any class and then transfers this information to the second system which calculates the probabilities for the predicted flare to be of C, M, or X class. ASAP reports probabilities every 15 minutes. This model is not run locally at CCMC; its output (figures and reports) is transferred from the University of Bradford.

\subsection{NOAA}
 
The NOAA probabilities (\url{http://www.swpc.noaa.gov/}) for solar flares are reported for the NOAA active regions once every 24 hours for prediction windows of 24, 48, and 72 hours. These probabilities are initially determined by a look-up table method that searches through catalogs of AR magnetic classification. In addition, forecasters consider the climatology, persistence, and their own expertise for reporting the flare probabilities \citep{Crown2012}.

\section{Linear Combination of Probabilities}

Let us call $f_{i,t+\Delta t}$ the probabilistic forecast made by the $i$-th method based on the information at time $t$ for the outcome between the times $t$ and $t+\Delta t$ \citep{Genre2013}. A linear combination reduces the information from $N$ forecasts to a combined value 

\begin{equation}
f_{t+\Delta t}^{c}(w)=\sum_{i=0}^{N} w_{i}\times f_{i,t+\Delta t}
\label{comb_prob_def}
\end{equation}

\noindent
which is a function of $w=\{w_{0},w_{1}, ..., w_{N}\}$, the set of combination weights. Combination weights represent the contribution of each method to the combined forecast. Their values range from 0 to 1.

The forecasts given by ASAP and NOAA provide probabilities for the occurrence of M-class (M1.0 - M9.9) and X-class (X1.0 and above) flares for a forecasting time window of 24 hours. ASSA provides differential forecasts, just like ASAP and NOAA, but for a prediction window of 12 hours. Therefore, in order to be included, the 12-h ASSA probabilities must be converted into 24-h probabilities. On the other hand, MAG4 prediction window is of 24 hours but its forecasts are cumulative, i.e. it predicts combined  M\&X flares and X-class flares alone, as well. In order to include MAG4 in the linear combination, its forecasts must be converted into differential forecasts (M- and X-class separately). See Appendix 1 for details on the conversion of ASSA and MAG4 forecasts. Thus the probabilities $\{P_{i}\}$ given by the ensemble of methods $\{\rm MAG4,ASSA,ASAP,NOAA\}$ can be combined, and Eq. \ref{comb_prob_def} takes the form

\begin{equation}
 P^{c}(w,t)=w_{\rm MAG4}P_{\rm MAG4}(t) + w_{\rm ASSA}P_{\rm ASSA}(t) + w_{\rm ASAP}P_{\rm ASAP}(t) + w_{\rm NOAA}P_{\rm NOAA}(t)
 \label{comb_probs_p}
\end{equation}

All probabilities in Eq. \ref{comb_probs_p} are given for $\Delta t=$ 24 hours, and therefore such label is omitted from this point on. For probabilistic forecasts, the combination weights must satisfy the condition $w_{\rm MAG4} + w_{\rm ASSA} + w_{\rm ASAP} + w_{\rm NOAA}=1$. Due to the similarities (physical foundation) shared by all four forecasting methods, some level of correlation is expected (and evidenced) between different probability time series. By performing this linear combination, we average out the disagreements among the individual methods and underline the common features that they display.

Linear combination in Eq. \ref{comb_prob_def} corresponds to a single-value combined probability for a the forecast time $t$. Eq. \ref{comb_probs_p}, on the other hand, defines a time series of the combined forecasts. In this study we will use time series of probabilistic forecasts corresponding to individual ARs in order to maintain the training and validation subsamples independent (Section 5) during the Monte-Carlo simulation \citep{Falconer2014}.

\section{Performance-based combination weights}

In order to calculate the set of weights that provide the optimal linear combination, we look into the performance history of each method in predicting past events \citep{Armstrong2001}. Our approach to solve this problem is to compare time series of probabilities $P(t)$ to the time series of events $E(t)$, occurring in a certain active region. Average values of $w_{i}$ over a sample of active regions should provide a measure of how much each method can be trusted in forecasting new events.

For a particular set of weights, the combined time series such as $P^{c}(t)$ from Eq. \ref{comb_probs_p} can be compared to the events time series $E(t)$ by applying a decision threshold $P_{th}$ to the former. This decision threshold is the value used to transform probabilistic forecast $P^{c}(t)$ into categorical (yes/no) forecast $F^{c}(t)$. Figure \ref{Comb_ts} displays an example of such thresholding process: the solid black horizontal line corresponds to $P_{th}=$ 25 \%. For $P^{c}(t)>P_{th}$, it corresponds to a "yes" forecast and it is assigned a value 100 in $F^{c}(t)$. For $P^{c}(t)<P_{th}$ a "no" forecast is obtained and therefore a zero value is assigned in $F^{c}(t)$. For measuring the similarity between $F^{c}(t)$ and $E(t)$, a relevant metric must be used. The performance of the ensemble forecast will depend on the choice of this metric.

 \begin{figure}
 \centering
 \noindent\includegraphics[width=40pc]{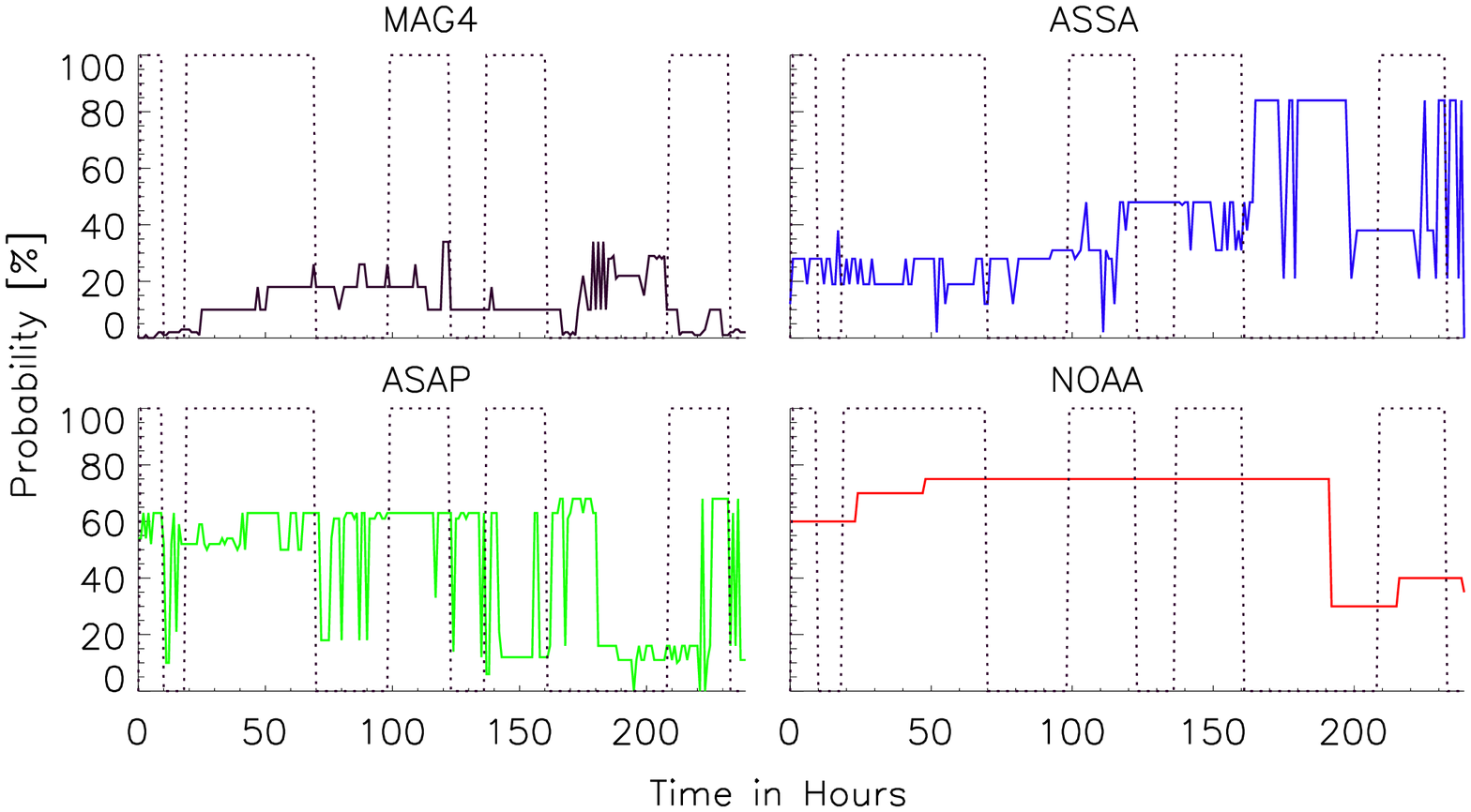}
 \caption{Times series of probabilities and events corresponding to AR 11429 for M-class flares. Solid lines correspond to probability time series while dotted lines corresponds to observations. Forecasting method probabilities are color coded as follow: black for MAG4, blue for ASSA, green for ASAP, and red for NOAA. It can be seen that all methods produce different probabilistic forecasts for the same photospheric conditions.}
 \label{time_series}
 \end{figure}

\subsection{Metric Optimization}

The difference between forecasts and observations can be quantified by an appropriate metric. Mathematically, two time series can be compared using a parameter such as the goodness-of-fit

\begin{displaymath}
\chi^{2} = \sum_{t}\frac{[E(t)-F^{c}(t)]^{2}}{\sigma^{2}}
\end{displaymath}

\noindent
where $\sigma^{2}$ is the variance of $E(t)$. However, even though $\chi^{2}$ is a standard measure of the error between prediction and events, it cannot determine the skill of the forecasting method since no comparison to a benchmark is done. A more relevant metric, commonly used for many types of forecasts, is the Heidke Skill Score \citep{Bloomfield2012}:

\begin{equation}
HSS = \frac{a+d-e}{n-e}.
\label{hss_1}
\end{equation}
All quantities in Eq. \ref{hss_1} are defined using the 2$\times$2 contingency table (see Table \ref{table2}). The contingency table counts how many times an event was both forecasted and observed (hits), forecasted but not observed (false positives), not forecasted but observed (misses), and neither forecasted nor observed (correct negative). The HSS metrics measures the performance of a forecasting method compared to a random prediction model used as a benchmark. HSS = 1 corresponds to a perfect forecast while HSS=0 corresponds to the opposite case in which the forecast has no skill compared to the benchmark and performs as a random guess. For consistency, comparison between the ensemble forecasts and other methods (e.g. the ensemble members) has been done in terms of the HSS metric.

In this investigation, we used our own routine for calculating HSS. Its proper functioning was verified using synthetic time series of probabilities and events. Time series had between 240 and 2400 total time steps. First, an event time series was created by randomly assigning values of 0 or 100 at each time step. If a particular time in the event time series has a zero value (no event), that same time in the probability time series has 0\%. For non-zero event times, a random probability between 1\% and 100\% is assigned. By constructing the time series in this particular way, we ensured that when calculating HSS we obtain HSS = 1 for $P_{th}$= 0\% since there is no 'false positive' or 'miss' cases.

Values of HSS depend on the set of combination weights $w$ which depends on the threshold value $P_{th}$, and therefore HSS = HSS$[w_{\rm MAG4}(P_{th}),w_{\rm ASSA}(P_{ th}),w_{ASAP}(P_{th}),w_{NOAA}(P_{th})]$. For each value of $P_{th}$, the most optimal combination of weights $w^{max}$ which maximizes the value of HSS has been determined by scanning the entire sub-space of $w$-values that satisfies the condition $w_{\rm MAG4} + w_{\rm ASSA} + w_{\rm ASAP} + w_{\rm NOAA}=1$. Values of the applied threshold have been varied from $P_{th}=$ 5\% to $P_{th}=$ 60\% with a step $\Delta P_{th}$ = 5\%.

 \begin{figure}
 \centering
 \noindent\includegraphics[width=30pc]{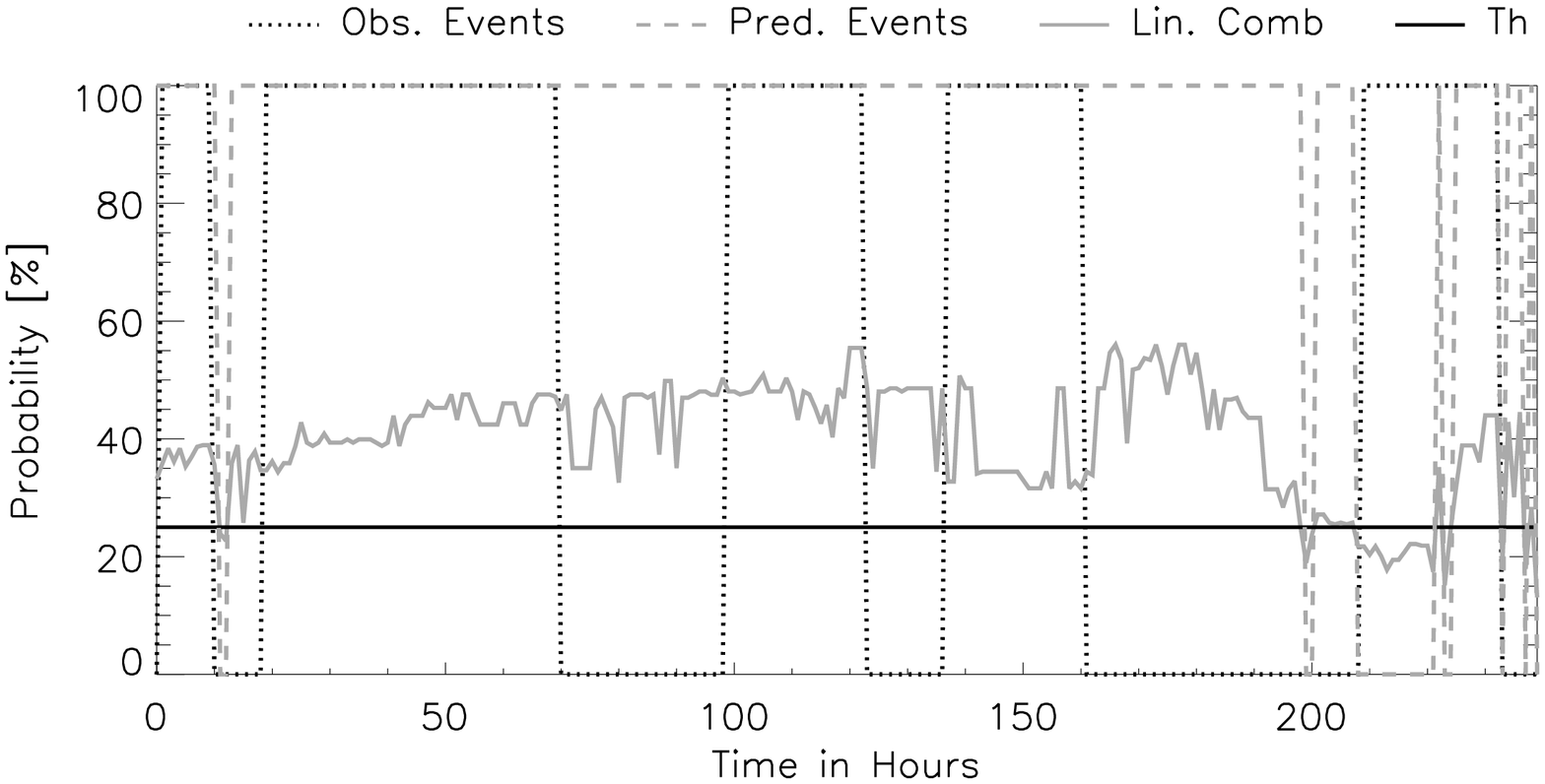}
 \caption{Example of the ensemble construction by the linear combination using the probability time series from AR 11429 (Figure \ref{time_series}). Individual methods time series were combined using Eq. \ref{comb_probs_p} and $w_{MAG4}=0.29$, $w_{ASSA}=0.17$, $w_{ASAP}=0.28$, $w_{NOAA}=0.26$ (solid grey curve). Linearly combined probabilistic forecast are transformed into categorical forecasts (dashed grey line) by applying a decision threshold (horizontal black line), e.g. $P_{th}=$ 25\%. Categorical forecast time series can be compared to the events time series by calculating the metrics based on the 2$\times$2 contingency table.}
 \label{Comb_ts}
 \end{figure}

\section{Validation}

Values of $w^{max}$ could, in principle, be different for each active region. Therefore, average values over multiple active regions are statistically more representative. Following \cite{Falconer2014}, in order to calculate the optimal combination weights, we implemented a Monte-Carlo (MC) algorithm that randomly selects 10 active regions from the studied sample (13 flaring active regions in total) and assigned them to the training subsample and the remaining three to the validation subsample. The random selection is done for the active regions: once a particular active is been assigned to one of the subsamples, the entire 10-hour time series remains in the subsample. This guarantees the independence of the two subsamples for statistical purposes. For each MC step, the algorithm calculates $w^{max}$ using the training subsample, and then it applies these weights to the validation subsample from which the 2$\times$2 contingency table and the performance metrics are calculated. Performance metric values depend on the parameter $P_{th}$ as well as $N_{\rm MC}$, the total number of MC steps. 

\begin{figure}
 \centering
 \noindent\includegraphics[width=40pc]{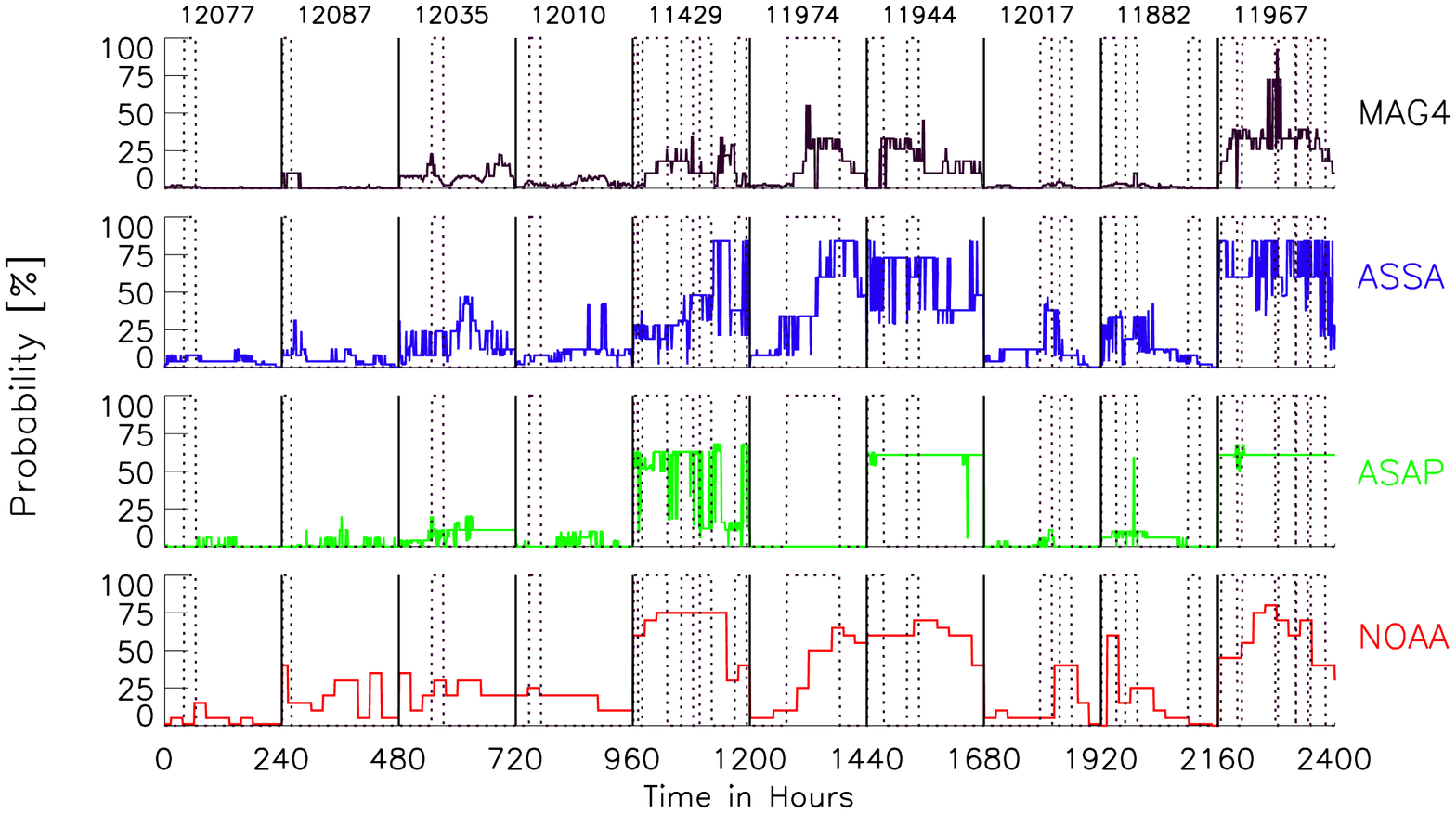}
 \caption{Extended times series of M-class flare probabilistic forecasts (solid curves) and events (dotted curves) constructed by putting together the times series of all ARs in the training subsample (10 ARs) for a particular Monte-Carlo step in the simulation. Colors identifying the methods are the same as in Figure \ref{time_series}. Five-digit numbers on the top of each plot are the AR identifiers.}
 \label{ext_ts}
 \end{figure}

\begin{figure}
 \centering
 \noindent\includegraphics[width=20pc]{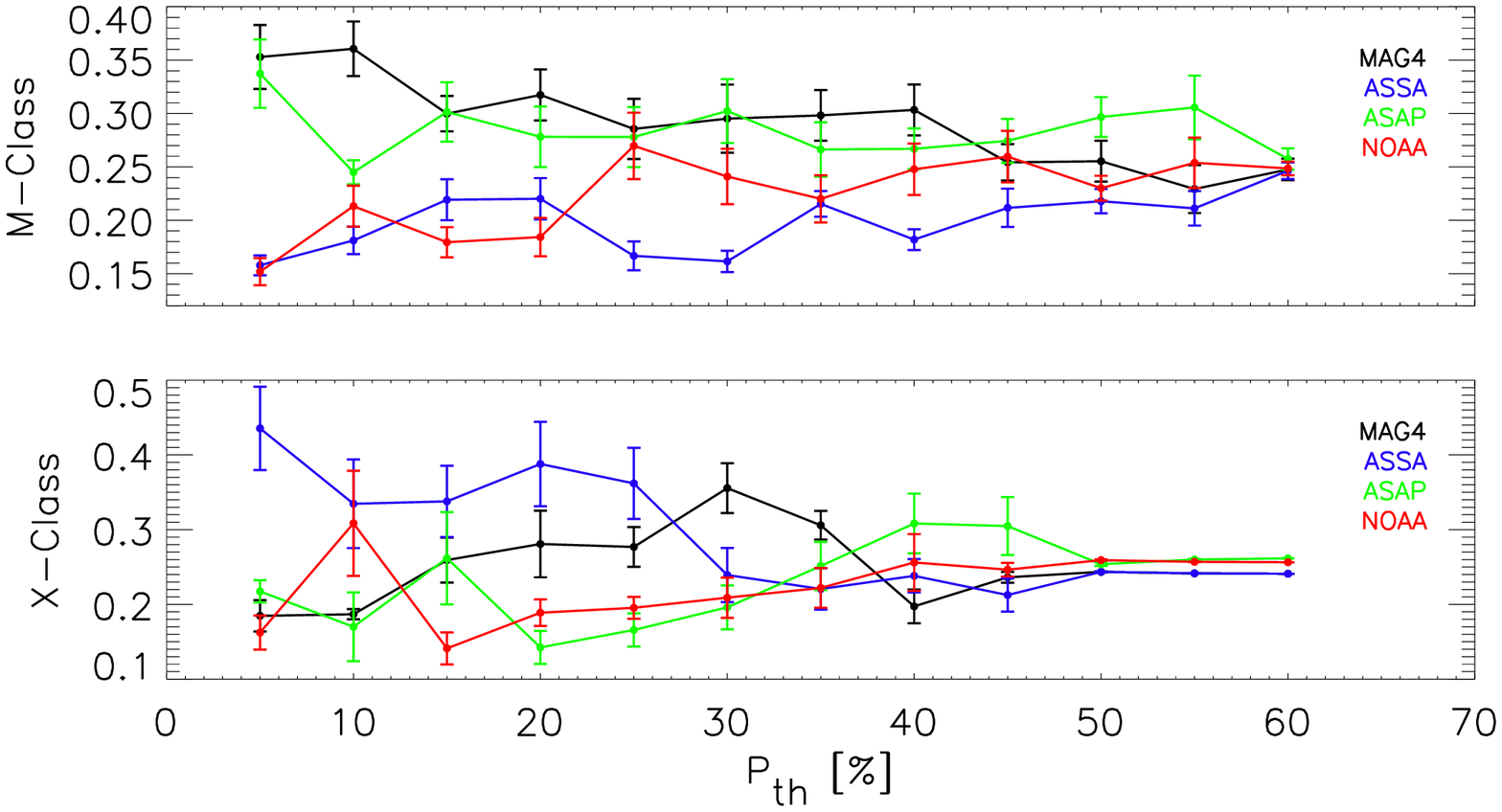}\includegraphics[width=20pc]{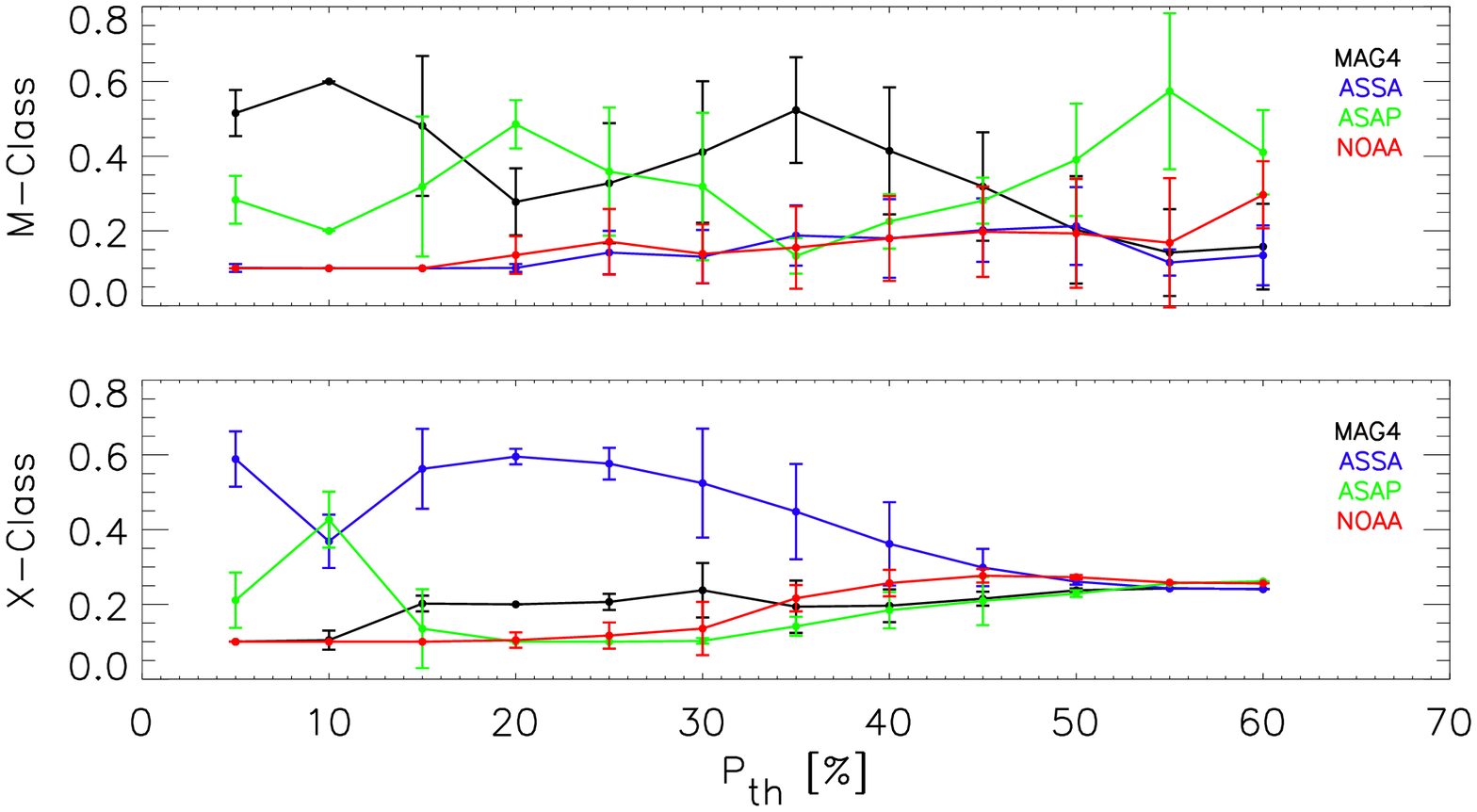}
 \caption{Combination weights for the ensemble forecasting method. Upper and lower panels display the weights for M-class flares and X-class flares, correspondingly, for each forecasting method as a function of the applied threshold $P_{th}$. Left panels correspond to the sample averaged values; right panels represent the optimization of the extended time series. In both cases, averages over all Monte-Carlo steps are shown. Uncertainties are given by the 1-$\sigma$ measures.}
 \label{comb_weights}
 \end{figure}

We calculated two sets of combination weights during the training process. For each MC step, we first analyzed each active region individually, calculating its set of weights $w^{max}_{i}$ as described in Section 4.1. Then, an average (AVE ensemble) over the training subsample was calculated, $w^{max}_{\rm AVE}=\langle w^{max}\rangle $. Secondly, we calculated another set of weights by applying the metric optimization procedure (Section 4.2) to the extended time series (ETS ensemble; see Figure \ref{ext_ts}) of probabilities $P^{c}_{\rm ETS}(t)$ and events $E_{\rm ETS}(t)$. The extended time series were constructed by joining time series of all active regions in the training subsample. By analyzing the extended time series, we calculated a set of weights $w^{max}_{\rm ETS} $ that maximizes HSS for all AR time series simultaneously but not necessarily individually. We expect $w^{max}_{\rm ETS}\ne w^{max}_{\rm AVE}$ since the HSS variates with the ratio of occurrence to non-occurrence forecasts \citep{JolliffeStephenson2003}. Figure \ref{comb_weights} displays both sets of combination weights for the largest MC step in our simulation.

Figure \ref{comb_weights} displays the values of combination weights obtained by averaging over the AR subsample $w^{max}_{\rm AVE}$ (left panels) and by applying the extended time series optimization $w^{max}_{\rm ETS}$ (right panels). It can be seen that measuring the combination weights using the two approaches results in very different values for the same forecasting method and the same threshold. For M-class flares, $w^{max}_{\rm AVE}$ ranges from 0.15 to 0.35 while $w^{max}_{\rm ETS}$ does from 0.1 to 0.6. For the X-class flares, the$w^{max}_{\rm AVE}$ range is 0.1 - 0.45 while the $w^{max}_{\rm ETS}$ range is 0.1 - 0.7. By examining both sets of weights, we have found that contributions from each method vary significantly with the value of $P_{th}$.

In addition to the HSS, there is a large number of metrics that can be derived from the 2$\times$2 contingency table (Table \ref{table2}) and be used to measure the performance of the ensemble forecast. We selected a group of metrics whose wide scope of applicability makes them appropriate for evaluating any forecasting method: the Percentage Correct (PC) score, the Probability Of Detection (POD) score, True Skill Score (TSS), and False Alarm Rate (FAR). Table 2 shows the definitions for each of these metrics. For this set of metrics, a perfect forecasting method should yield PC$=$POD$=$HSS$=$TSS $=$1 and FAR $=$0.

\subsection{Algorithm-generated versus Human-adjusted probabilities}

The process by which the probabilistic forecasts given by NOAA are determined is very different from that of the three other methods since includes human judgement and expertise. This difference suggests the idea of constructing an ensemble method by only combining the forecasts with similar techniques: MAG4, ASSA, and ASAP -- the fully automated methods -- and then comparing this ensemble against the ensemble that includes NOAA and the NOAA method alone. Thus, we repeated the same training-validation process, this time excluding the NOAA predictions from the ensemble.

\begin{figure}
 \centering
 \noindent\includegraphics[width=18pc]{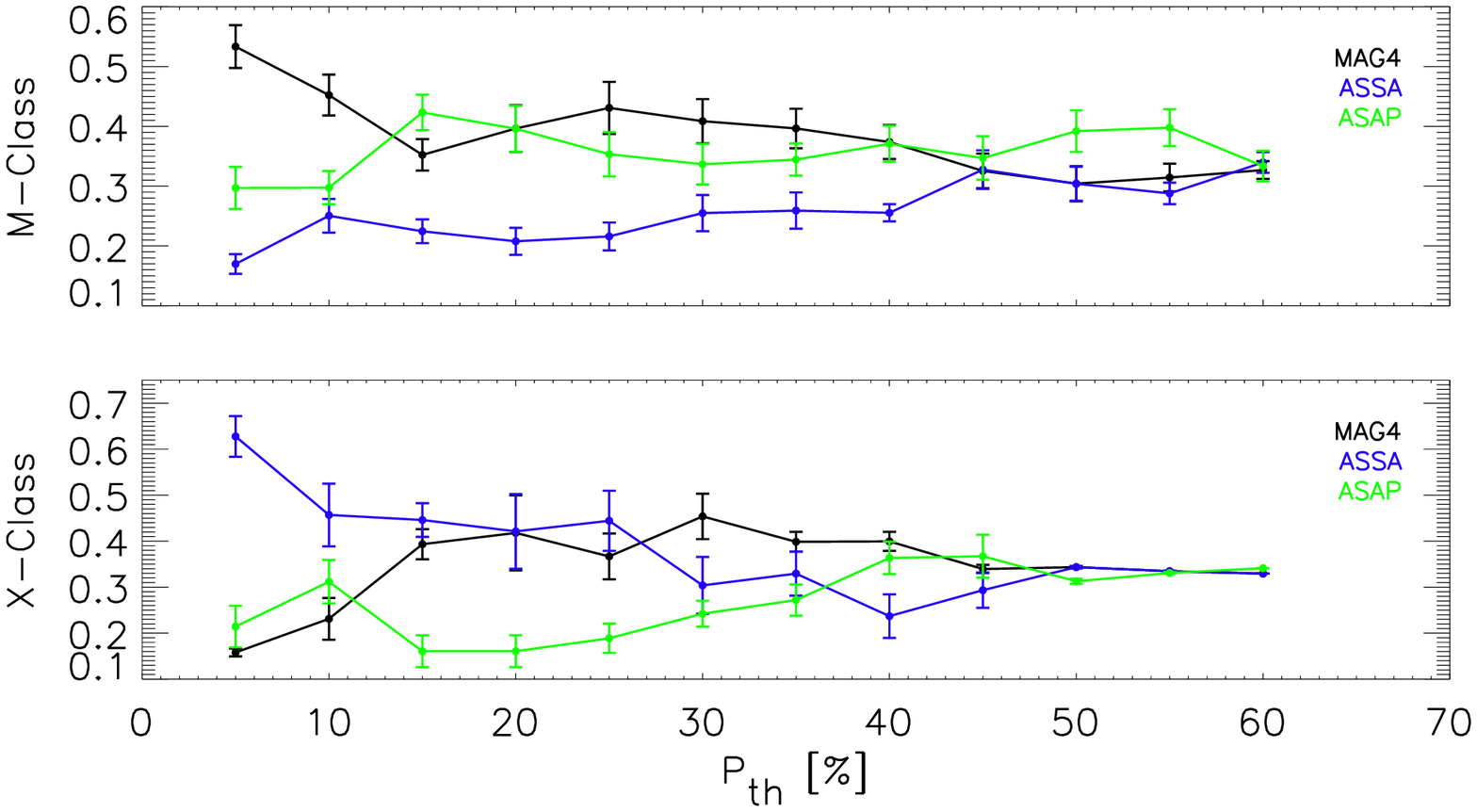} \includegraphics[width=18pc]{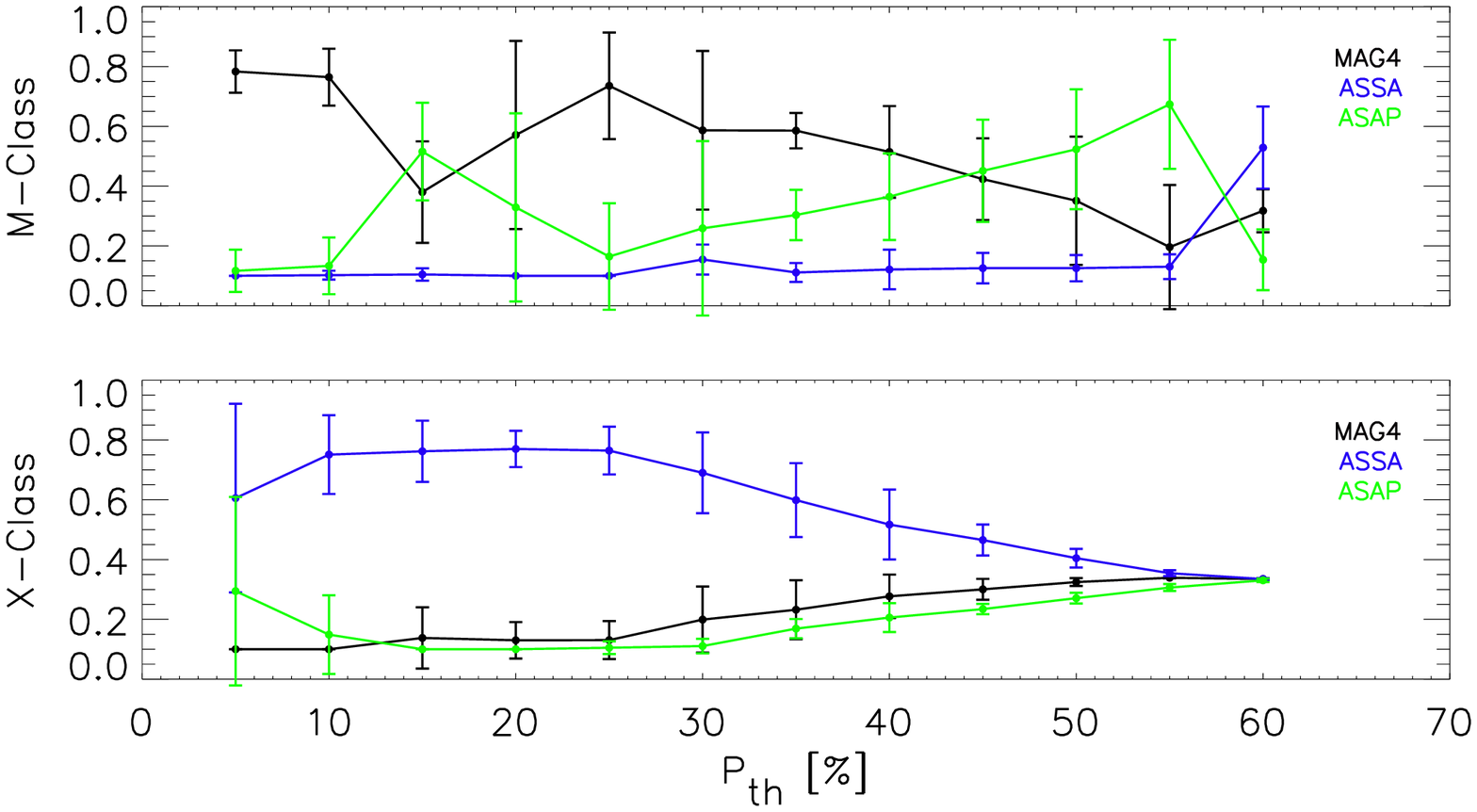}
 \caption{Combination weights $w$ for the ensemble forecast built using only fully-automated forecasting methods: MAG4 (black), ASSA (blue), and ASAP (green). Each panel displays the variation $w$ with the applied threshold $P_{th}$. Left panels correspond to the weights calculated by AR-sample averaging; right panels show the weights obtained by the optimization of the extended time series.}
 \label{weights_th_3m_methods}
 \end{figure}

Figure \ref{weights_th_3m_methods} displays the weights $w^{max}_{\rm AVE}$ (left panel) and $w^{max}_{\rm ETS}$ (right panel) calculated using only MAG4, ASSA, and ASAP in the ensemble. Values of the combination weights are expected to be different from those in Figure \ref{comb_weights}, because of the normalization constrain discussed in Section 3. However,  their behaviors with the $P_{ht}$ are similar to those in the NOAA-including case: no method completely dominates the ensemble for the entire range of applied threshold.

\section{Results and Discussion}

\begin{figure}
 \centering
 \noindent\includegraphics[width=25pc]{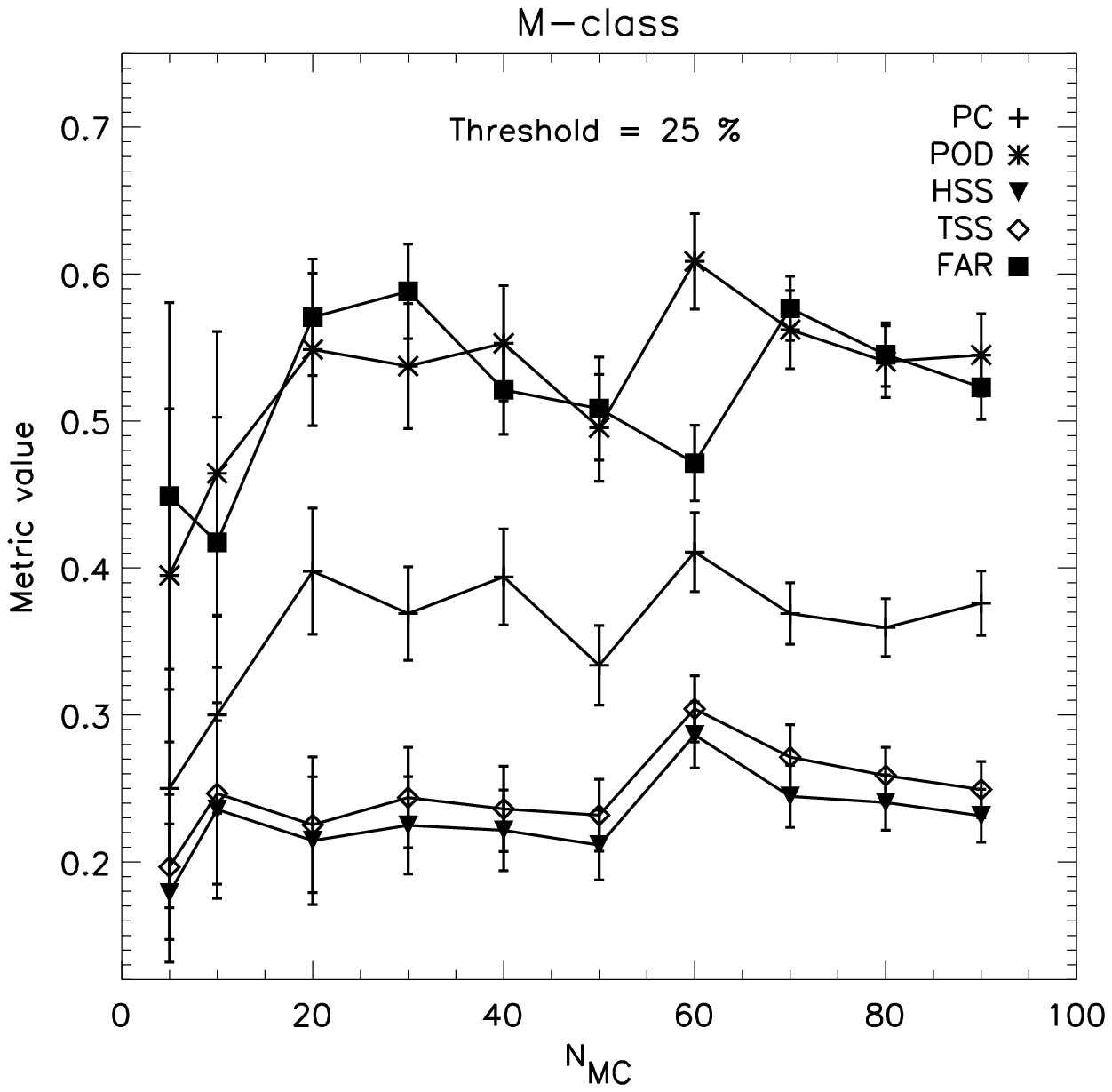}\\
 \caption{Variability of averaged values of the contingency table statistics (PC, POD, HSS, TSS, and FAR) with the total number of Monte Carlo steps for a particular value of threshold, $P_{th}=$ 25 \%. For $N_{\rm MC}>$ 60,  fluctuations in the measured metrics are less observed and therefore more stable values achieved. This stabilization is due to oversampling of the set of AR used in this investigation. Error bars correspond to the $\sigma/\sqrt{N_{\rm MC}}$ values.}
 \label{metrics_n}
 \end{figure}

The training-validation process described in Section 5 was performed for several numbers of total MC steps, $N_{\rm MC}$, after which all the quantities (both metrics and weights) were averaged over $N_{\rm MC}$. Figure \ref{metrics_n} displays the dependence of the contingency table metrics with $N_{\rm MC}$ for M-class flares and $P_{th} = 25\%$. All five metrics seem to show a weak dependence with $N_{\rm MC}$. For $N_{\rm MC}>$ 60 all metrics seem to approach their stable values, showing little fluctuations because of the oversampling of the AR set -- that is, all possible combinations of randomly-produced training and validation subsets are accounted for and more realizations do not contribute to the MC averages. In this section, all metrics values correspond to the largest MC total steps, $N_{\rm MC}$=90. Uncertainties in the measures are given by $\sigma/\sqrt{N_{\rm MC}}$, where $\sigma$ is the standard deviation of the distribution characterized by $N_{\rm MC}$.

\begin{figure}
 \centering
 \noindent\includegraphics[width=25pc]{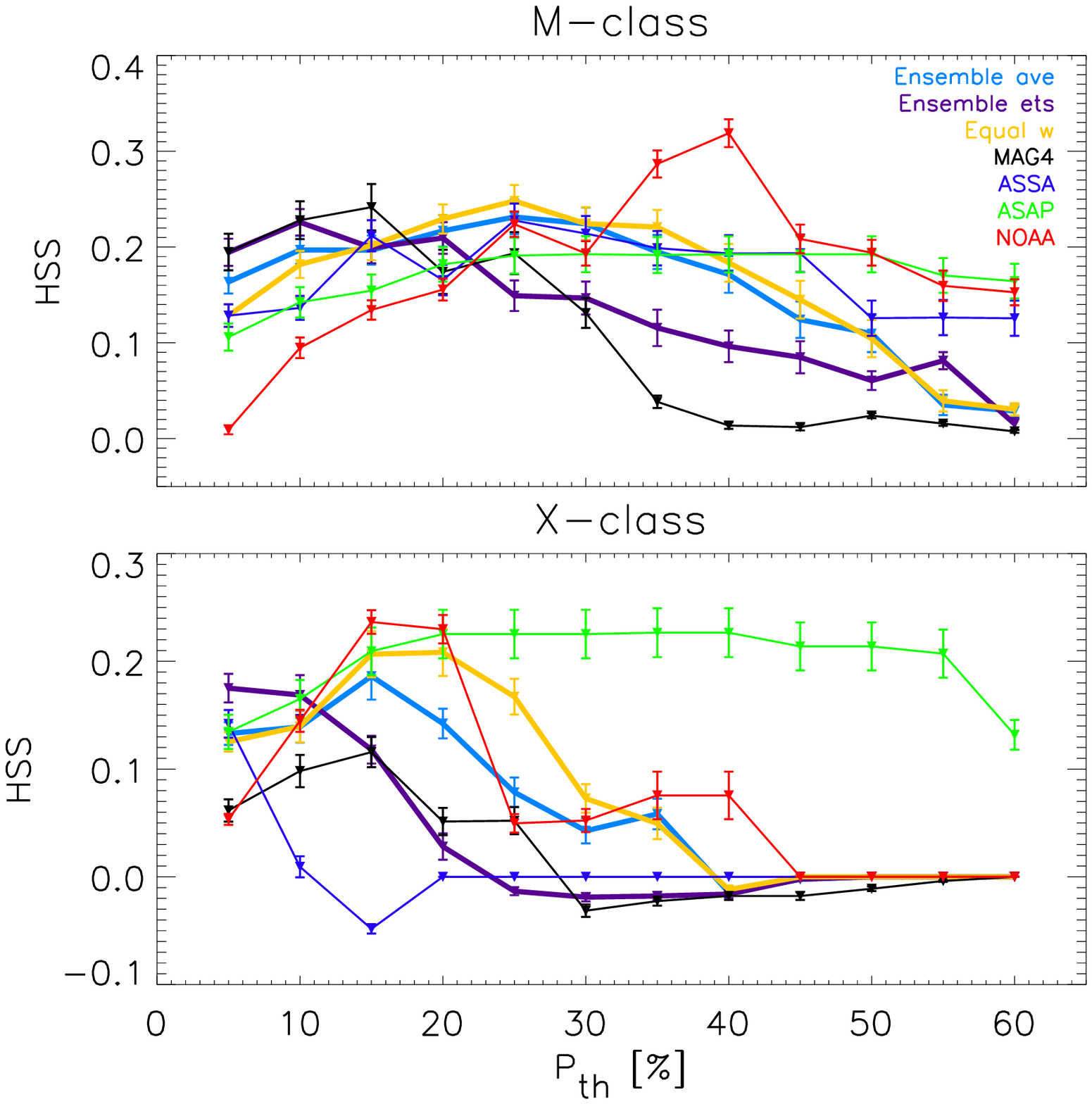}
 \caption{Heidke Skill Score (HSS) as a function of the applied threshold $P_{th}$ for the ensemble forecast and the ensemble's members. Ensemble models are shown with thick lines in both panels: light blue for performance-based averages weights ensemble, purple for the extended time series optimization, and yellow for equally-weighted ensemble. Thin curves correspond to the ensemble members: MAG4 (black), ASSA (blue), ASAP (green), and NOAA (red). For M-class flares (top panel) ensemble forecast reaches maximum value HSS $\approx$ 0.27 at $P_{th}=$ 25 \% for the equally-weighted model. In the X-class case, the equally-weighted model display HSS $\approx$ 0.21 for $P_{th}=$ 15 \%. }
 \label{hss_th}
 \end{figure}

In order to compare the performances of the ensemble forecast and its members, during the validation phase in our MC algorithm we also calculated metrics for all ensemble member methods as well as for the equally-weighted ensemble, for which $w_{i}=0.25$. In Figure \ref{hss_th} we compare the HSS curves for M-class flares (upper) and X-class flares (lower). By inspecting these plots (Figure \ref{hss_th}) it is possible to determine the conditions (threshold and weight values) for which the ensemble categorical forecast displays higher values of HSS compared to all the ensemble's members.

\begin{figure}
 \centering
 \noindent\includegraphics[width=25pc]{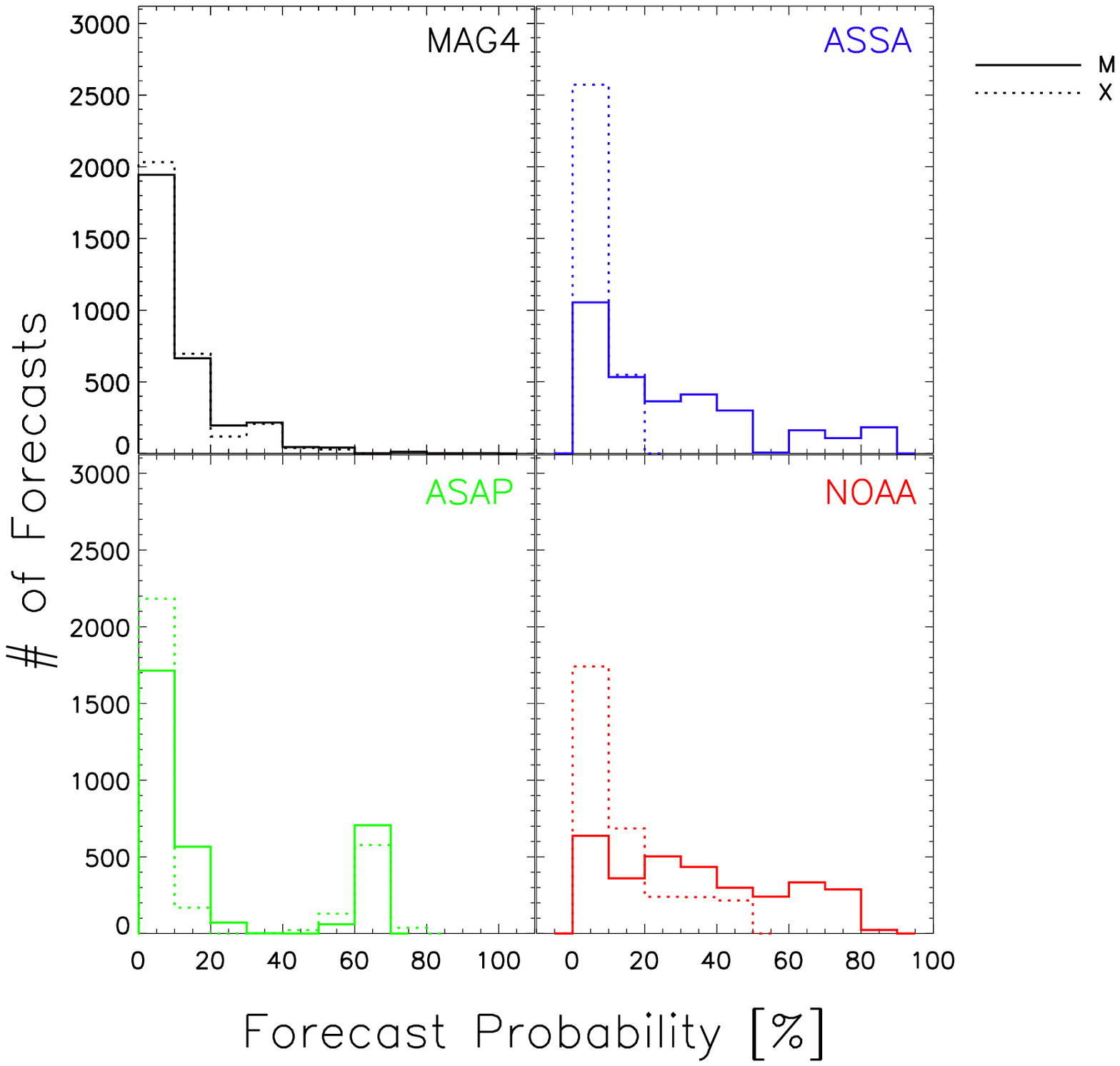}
 \caption{Distribution of probabilistic forecasts for the ensemble members for both studied classes of flares. Histograms were calculated using a bin size of 10 \%. For the considered statistical sample of active regions investigated here, fully-automated methods (MAG4, ASSA, and ASAP) produce distributions mostly concentrated in the lowest bins, $P<$ 20 \% for both flare classes. Compared to other methods, NOAA produces a flatter distribution of probabilities for M-class flares. A more representative sample would be needed to validate these tendencies.}
 \label{histograms}
 \end{figure}

From Figure \ref{hss_th} we observe that for M-class flares, all methods seem to describe a similar tendency: HSS values start from a lower value, reach a maximum for a specific values of $P_{th}$ and then decreases. All methods show HSS $>$ 0.1 for $P_{th}=$5 \%, the lowest threshold value, except NOAA that shows HSS $\approx$ 0.0. Most ensemble models and individual members reached maxima (HSS $\approx$ 0.22 - 0.25) for $P_{th}=$ 10 - 25 \%, with the equal weights model having the highest value in this range. On the other hand, NOAA shows its maximum HSS $\approx$ 0.35 for $P_{th}=$ 40\%, which is the highest HSS value for all the studied methods (ensemble and individual methods). The difference between model and NOAA predictions can be understood by inspecting the histograms in Figure \ref{histograms}. The NOAA method systematically provides higher probabilities for M-class flares compared to ASSA, ASAP, and MAG4. This bias in the distribution of probabilities could be attributed to the human factor involved in determining the probabilistic forecast. For X-class flares, most methods display values of HSS $>$ 0 for $P_{th}<$ 45 \%, but the decay of the HSS values with $P_{th}$ is faster than that for the M-class flares. For X-class events, the equally-weighted ensemble (yellow) and the average ensemble (light blue) curves are less similar to each other, while the M-class flares show the opposite tendency. The equally-weighted ensemble model provides HSS $\approx 0.20$ for $P_{th}=15$ \% (the highest values for an ensemble model) although the NOAA method displays HSS $\approx 0.24$. The apparent poor performance of most methods (including the ensembles) can be caused by the small number of X-class events included in our AR sample. However, it seems like the ensemble forecast is able to improve the accuracy of the prediction in spite of the limited statistics.

In terms of the HSS values, forecasts from an ensemble method are accurate the most for $P_{th}=$ 25 \% for M-class flares and $P_{th}=$ 15 \% for X-class flares. In the M-class case, the AVE ensemble displays values of HSS very close to those of the equally-weighted ensemble method, which can be verified by inspecting the values of $w^{max}_{\rm AVE}$ in Figure \ref{comb_weights}.

\begin{figure}
 \centering
  \includegraphics[width=18pc]{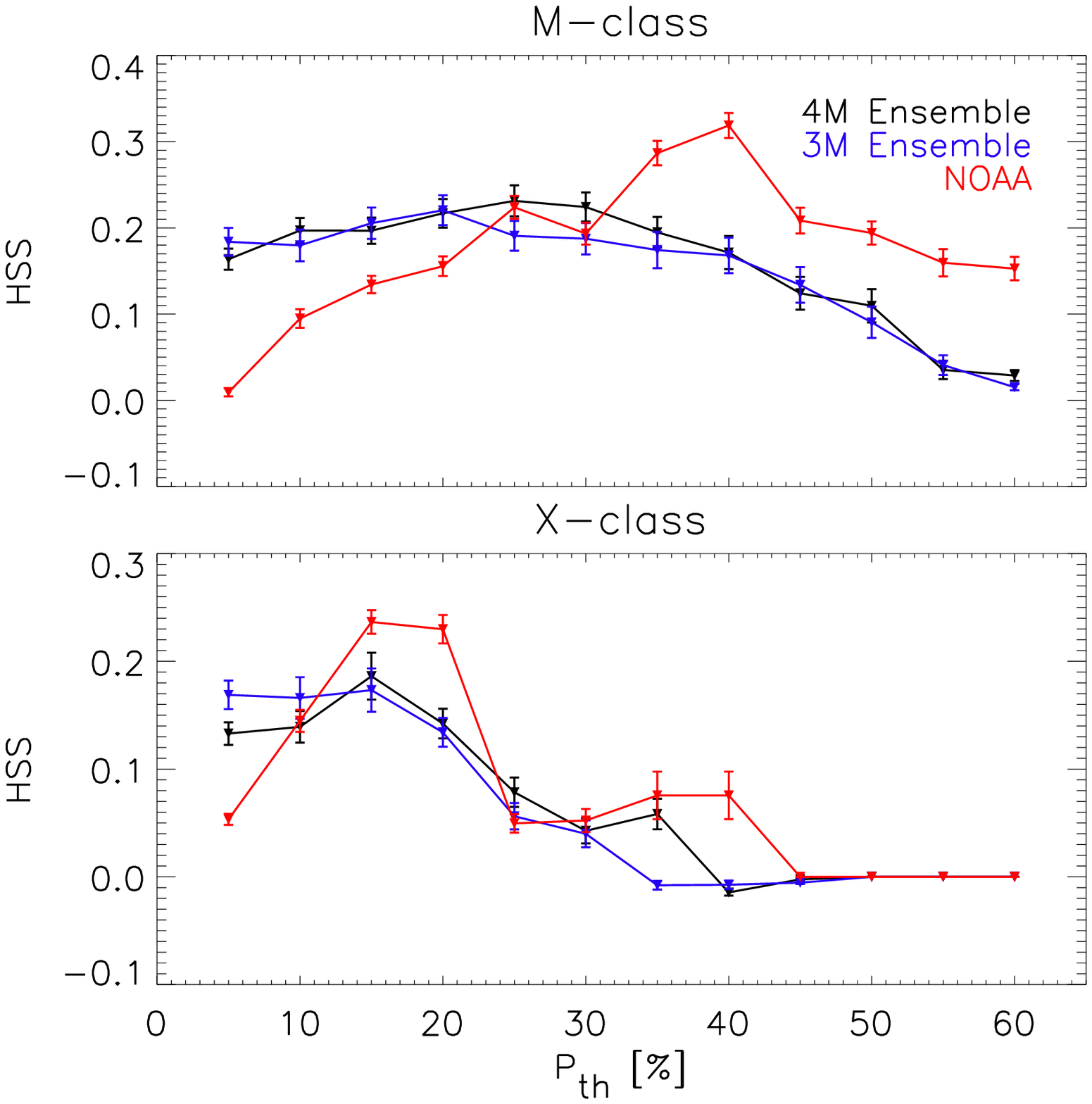}\includegraphics[width=18pc]{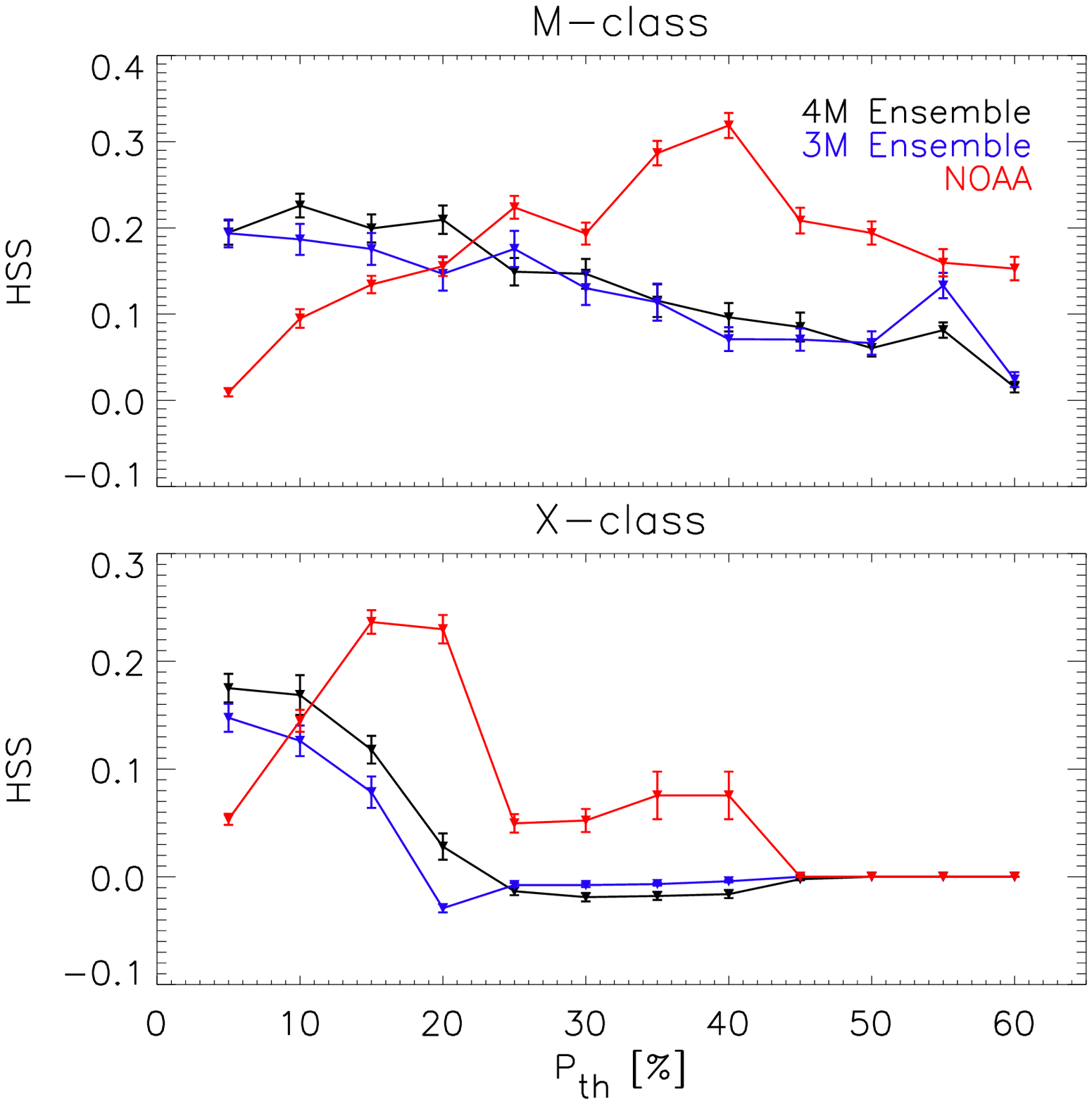}
 \caption{Values of HSS($P_{th}$) for M-class (upper panels) and X-class flares (lower panels) calculated with performance-based weights: AR sample average (left) and optimization of the extended time series (right). In all panels comparisons are made between the ensemble model that includes all four methods (black), the ensemble including only fully-automated forecasting methods (blue), and the human-influenced forecasts from NOAA (red). The inclusion of the NOAA forecasts in the ensemble prediction generally results in an improvement of the HSS performance metric.}
 \label{metrics_th_3m_methods}
 \end{figure}

In Figure \ref{metrics_th_3m_methods} we compared the HSS values for the ensemble that included (black) and excluded NOAA (blue) as a member. For M-class flares (upper panels) the inclusion of the NOAA method in the AVE ensemble seems to increase the HSS values for $P_{th}=$ 25 - 35 \% while does the same for $P_{th}=$ 10 - 20 \% for the ETS ensemble. For X-class flares, excluding NOAA seems to make the average ensemble to perform better for $P_{th} < $ 10 \%, while including it increases HSS values for the ETS ensemble for a wider range of thresholds ($P_{th} <$ 25 \%). As measured by HSS, combining only fully-automated forecasting methods to construct the ensemble does not seem to provide a much better prediction. On the contrary, including the human-generated forecasts from NOAA leads to a noticeably better performing ensemble prediction.

The better performance of the NOAA method over the ensembles observed in Figure \ref{metrics_th_3m_methods} can be attributed to the forecaster judgement. It is clear that fully-automated forecast methods have room for improvement before they can hope to ``replace the human in the loop". While it is reasonable to expect that NOAA prediction performance remains with larger active regions sample, we also expect improvement of ensemble predictions via improved combination weights.

\begin{figure}
 \centering
 \noindent\includegraphics[width=30pc]{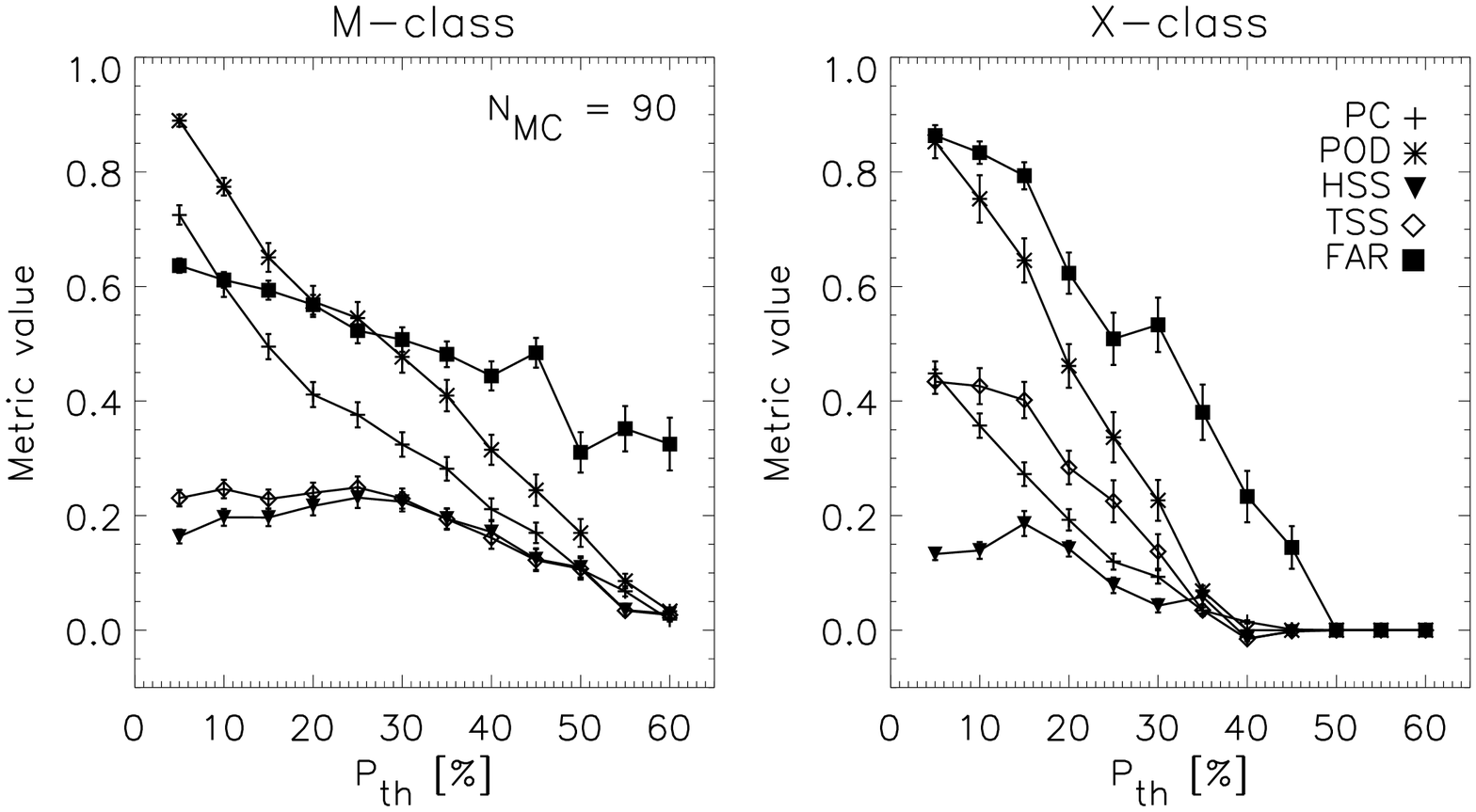}
 \caption{Performance metrics curves for the sample averaged ensemble. In each flare class -- M (left) and X (right) -- the percentage correct (PC), probability of detection (POD), Heidke skill score (HSS), true skill score (TSS), and false alarm rate (FAR) are shown. All non-skill metrics (PC, POD, and FAR) reach their maxima for the lowest  threshold value ($P_{th}^{min}=$ 5\%) and decrease with increasing the threshold. HSS and TSS metrics demonstrate maximum values for $P_{th}\ne P_{th}^{min}$. Less accurate metric values for the X-class flares are likely associated with the insufficient number of events present in the sample. }
 \label{metrics_th}
 \end{figure}

Figure \ref{metrics_th} displays the curves corresponding to the performance metrics for M and X classes using $w^{max}_{\rm AVE}$. For both cases shown in Figure \ref{metrics_th} it can be seen that PC, POD, and FAR decrease with the increasing threshold, displaying the maximum values for the lowest $P_{th}$. This behavior seems natural since increasing threshold translate to decreasing the hits and false positives (a and b in Table \ref{table2}) and increasing misses and correct negatives (c and d). On the other hand, the skill scores (HSS and TSS) reach their maxima around 20 - 25 \% and 15 \% for M- and X-class flares, correspondingly. These are the values of $P_{th}$ for which the ensemble forecast can be considered optimal.

For the X-class flares case, performance metrics curves do not look as smooth as in the M-class case, and their statistical uncertainties appear larger as well. The poorer performance obtained in the X-class case should be attributed to fewer active regions with X-class events included in our sample.

\subsection{Attributes of the Ensemble Forecast}

The construction of the ensemble method by linearly combining several forecasts improved not only the categorical (yes/no) forecast  in terms of the HSS metrics but also the probabilistic forecast itself. The key attributes of a probabilistic forecasting method are {\it accuracy, reliability,} and {\it resolution} \citep{Balch2008}. The accuracy of a forecasting system can be described by the Brier Score defined as

\begin{equation}
QR=\frac{1}{N_{\rm P}}\sum_{i=1}^{N_{\rm P}}(P_{i}-E_{i})^{2},
\label{qr}
\end{equation}
\noindent
where $N_{P}$ is the total number of predictions ($P_{i}$) given by the model and $E_{i}$ are the observations. Eq. \ref{qr} quantifies the average degree of correspondence between individual forecasts and observations. The reliability relates the event frequency of occurrence (number of events forecasted with $P_{i}$/number of forecasts with $P_{i}$ value) to the predicted probability

\begin{equation}
REL=\frac{1}{N_{\rm P}}\sum_{i=1}^{T}N_{\rm P_{i}}(\langle P_{i}\rangle-\langle E_{i}\rangle)^{2},
\label{rel}
\end{equation}
\noindent
while the resolution is the ability of a forecasting method to recognize {\it a priori} an event occurring with a probability which is different from the climatology value \citep{JolliffeStephenson2003}:

\begin{equation}
RES=\frac{1}{N_{\rm P}}\sum_{i=1}^{T}N_{\rm P_{i}}(\langle E_{i}\rangle-\overline{E})^{2}.
\label{res}
\end{equation}

\begin{figure}
 \centering
  \includegraphics[width=18pc]{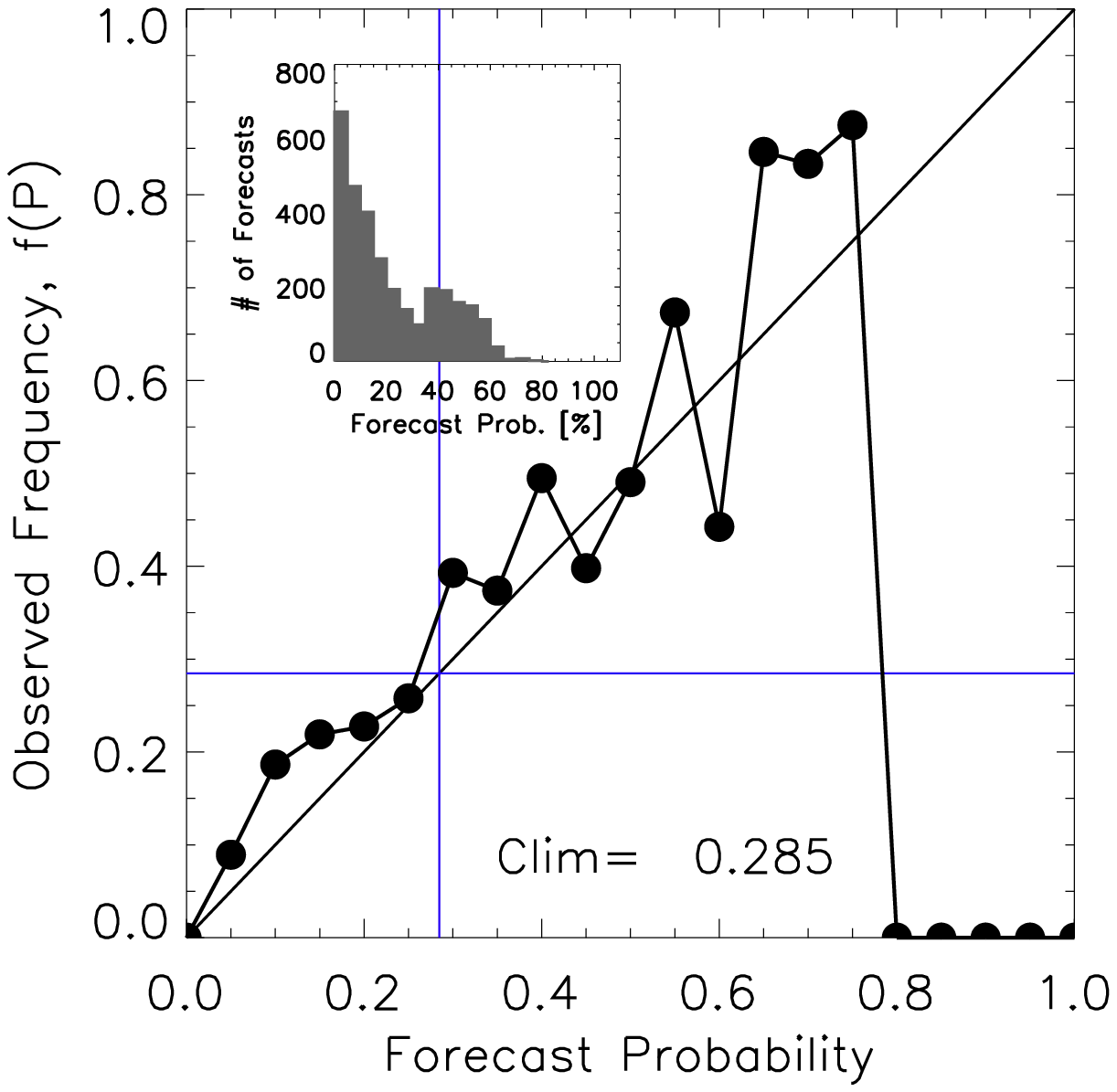}\includegraphics[width=18pc]{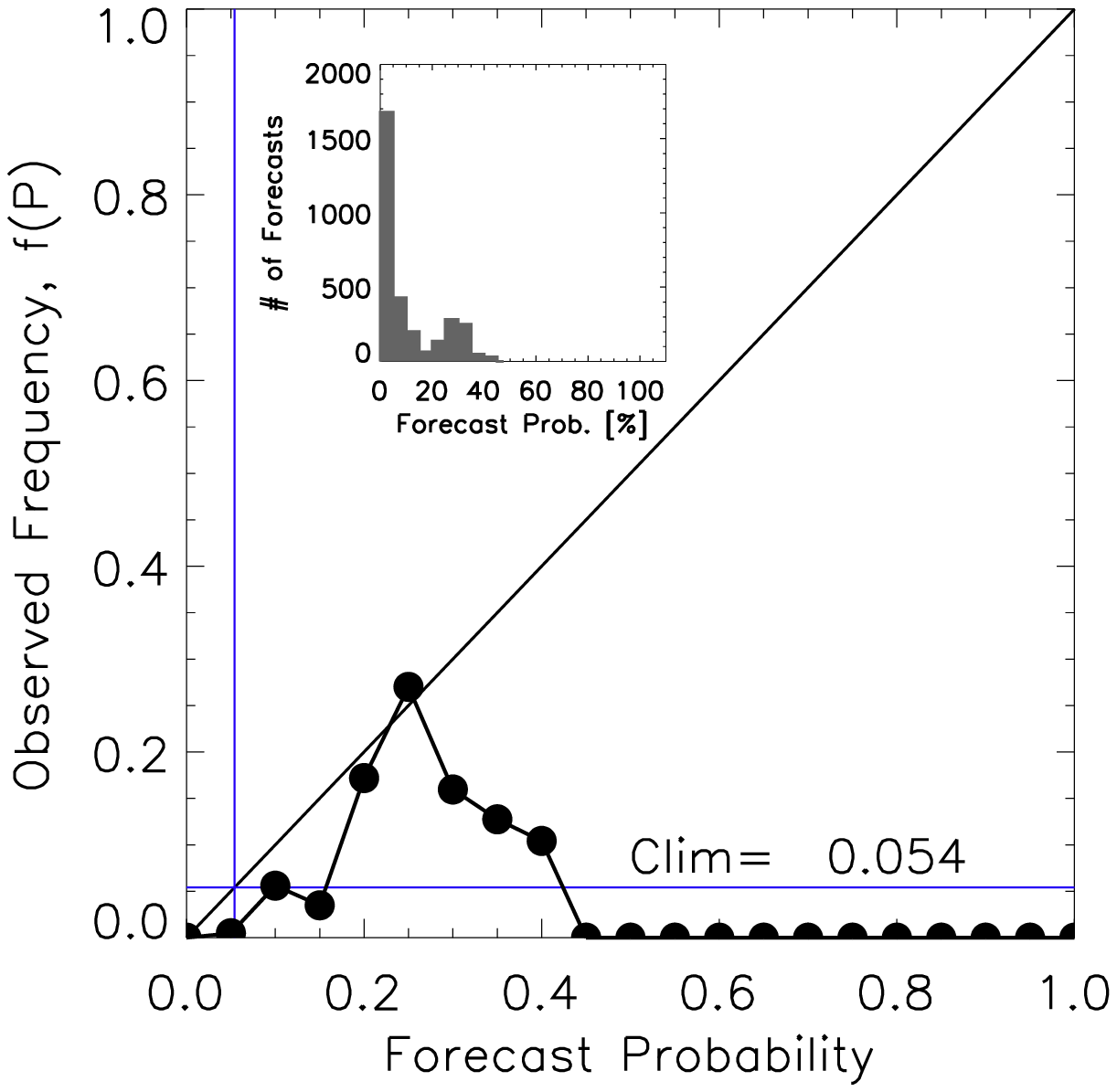}
 \caption{Attributes diagrams for the equally-weighted ensemble forecast of the M-class (left) and X-class (right) flares. Black circles correspond to the observed relative frequency of events for a given probability range. The diagonal solid line represents the zero reliability case. Horizontal and vertical lines mark the climatology value for each flare category. In the subset of M-class, the reliability curve follows the diagonal, which indicates that a linearly combined ensemble model improves the reliability of the probabilistic forecast.}
 \label{rel_plots}
 \end{figure}

In Eqs. \ref{rel} and \ref{res}, $T$ corresponds to the number of probability ranges (bins; see Figure \ref{rel_plots}), $N_{\rm P_{i}}$ is the number of ``model forecasts'', $\langle P_{i}\rangle$ is the mean forecast probability, and $\langle E_{i}\rangle$ is the event occurrence frequency for each range, while $\overline{E}$ is the climatology value of our sample.

The resolution of a forecasting method is the ability of the method to adjust its predictions for a particular level of activity. For example, let us assume that a flare forecasting method predicts an X-class flare using as probability the statistical occurrence rate for that flare category over one year during solar maximum. For such climatology forecast, we would expect high reliability: the flare occurrence frequency is consistent with the forecast probability. However, this climatology forecast does not have the means of distinguishing the conditions when the X-class flare will occur with a higher or lower probability. The higher the resolution, the better the forecasting method identifies these deviations from the climatology value.

For completeness, we also calculate a Skill Score (SS) for the probabilistic forecast. Using the accuracy value, $QR$, we calculate

\begin{equation}
SS = \frac{\overline{QR}-QR}{\overline{QR}}
\end{equation}

\noindent
where $\overline{QR}$ is the accuracy (Eq. \ref{qr}) when the climatology is used as the predictor.

Values of accuracy, reliability, and resolution range from 0 to 1. In terms of these attributes, $QR$ and $REL$ values closer to 0 correspond to a better accuracy and reliability, correspondingly, while values of $RES$ closer to 1 indicate a better resolution. These conditions (small accuracy, small reliability, and large resolution) also reduce the overall mean square error of the forecast \citep{Balch2008}. Values of $SS$ range from -1 to 1 (perfect skill) with 0 indicating that the probabilistic forecast has no skill over the climatological predictor.

Table \ref{table3} shows the values of $\langle P_{i}\rangle$, $QR$, $REL$, $RES$, and $SS$ for the three ensemble models we have constructed ($\rm AVE$, $\rm ETS$, and equal weights models) as well as for the four ensemble members (MAG4, ASSA, ASAP, NOAA). For M-class flares, the ensemble improves $QR$ and $REL$ values compared to individual members. In this case, the optimal accuracy and reliability are obtained when the ensemble is constructed using equal weights. The resolution (RES), on the other hand, is not improved by the ensemble in its present form. As seen from the $SS$ values in Table \ref{table3}, for M-class flares, all ensemble models improved the skill compared to individual members. In the case of the X-class flares, the ensemble combination does not provide the most optimal values for the attributes ($QR$, $REL$, $RES$) and skill ($SS$) compared to all individual members; again, the statistics are likely hampered by the small number of X-class events in the studied sample.

The skill score (SS) reported in Table \ref{table3} must be carefully interpreted. Values of SS $\sim$0.1 may appear as marginal improvement over the climatological predictor. However, skill scores obtained here are comparable to those obtained in \cite{BarnesLeka2008} and \cite{Crown2012}. It is important to keep in mind the differences in defining the events between those investigations and the ensemble method here presented (see Appendix A for the definition of events). We believe that our values of SS (and other metrics) are affected by the selection of active regions in our sample and that using a much bigger sample could improve the skill score values since more accurate combination weights can be calculated. However, for the purposes of our investigation, the SS values demonstrate the effectiveness of the ensemble method.

In addition to this, Figure \ref{rel_plots} displays the attribute diagrams for the equally-weighted ensemble model for M- and X-class flares. The attributes diagram is a visual interpretation of the information contained in Table \ref{table3}. It shows the relative occurrence frequency $f(P_{i})$ of the observed events (flares) for the cases when the events were forecasted to occur with the probability $P_{i}$ \citep{Wilks2006,JolliffeStephenson2003}. In the diagram, the climatology frequency is indicated by the horizontal and vertical blue lines. The horizontal line is known as the ``no resolution line". The inset in both plots of Figure \ref{rel_plots} shows the resulting histograms of probabilities describing the constructed ensemble. The information contained in the reliability curve can be used to correct forecasting errors due to biases. 

A reliable forecast shows $f(P_{i})=P_{i}$ for all values of $P_{i}$. In Figure \ref{rel_plots} (left) for M-class flares, the ensemble forecast shows a tendency that roughly follows $f(P_{i})=P_{i}$. For those values of $P_{i}$ where $f(P_{i})$ is above the diagonal, the ensemble forecast is underpredicting the events. On the contrary, if $f(P_{i})$ is less than $P_{i}$, the ensemble forecast is overpredicting the events. As evidenced by Table \ref{table3}, using the combination weights that yielded the maximum HSS values (AVE model, Figure \ref{comb_weights}; $P_{th}=$25 \%) for constructing the reliability plot produces a curve with more pronounced deviations from the diagonal and an overall reliability lower compared to that obtained using equal weights. 

For the X-class flares, improvement of the probabilistic forecast by using the ensemble model seems less obvious, but it could likely be achieved with increasing the number of events in the statistical sample.

\section{Summary}

We have demonstrated how an ensemble predicting model can be constructed using a group of several forecasting methods. This investigation is the first effort toward developing an ensemble-based prototype system for flare forecasting. In constructing our prototype, we learned several valuable lessons. First, an ensemble prediction having potential to outperform individual members can be constructed using a simple linear combination method. Second, it is important to utilize appropriate metrics in training and validating the ensemble method. And third, the human factor is still valuable in making flare forecasts.

In this initial work, the ensemble method provided the most accurate probabilistic prediction and the best categorical prediction for certain probability-threshold values. We have constructed an optimized forecasting ensemble by using linear combinations of member methods with different sets of combination weights: 1) performance-based weights, by analyzing individual active region (AR) time series and then averaging over the sample, 2) peformance-based weights, by analyzing the extended time series with all ARs included, and 3) equal weights. We evaluated the performance of the ensemble models by performing a validation process that included evaluating the contingency table statistics of the categorical forecast as well as the attributes and skill of the probabilistic forecast.

For the categorical prediction, our results showed that the sample-averaged weights make the ensemble model perform similarly to the equally-weighted model. Maximum values of the HSS are achieved when a threshold ($P_{th}$) of 25 \% for M-class flares and 15 \% for X-class flares is used for converting probabilistic forecasts into categorical forecasts. For these values of  thresholds, the ensemble forecast should be preferred over any of the ensemble members. This particular threshold-dependent behavior can be attributed to biases in each method when determining the probabilities -- these biases are evident in the individual methods' distribution of probabilities. In addition, constructing the ensemble forecast by combining only fully automatic forecasts produces a different set of combination weights and results in lower (compared to the ensemble that included human-generated NOAA predictions) values of HSS for some ranges of $P_{th}$.

On the other hand, the attributes and skill of the probabilistic forecast suggested that an ensemble model can provide better predictions than any of its members. In particular, for the M-class flare ensemble prediction, two out of three key attributes (accuracy and reliability) along with the skill of the probabilistic ensemble forecast were improved over the values yielded by the ensemble members.

While the performance metrics indicated the substantial challenge in predicting major flares, we believe that probabilistic and categorical forecasts of the ensemble model can be further improved by improving the active region statistics used in building the model. It is necessary to revalidate this analysis using a larger sample of active regions exhibiting considerably different levels of flaring activity. Once the statistics of the analysis have been improved, we will implement the ensemble forecasting of flares in a real-time prototyping environment. This will be done in a future study.


%
\appendix 
%
\section{Time Series Construction}

We constructed the probabilities time series $P_{i}(t)$ for each flare class using the hourly probabilities that each method reports for a particular active region. Time series of events $E(t)$ are constructed using the NOAA Solar Events Reports, in which detected flares are reported according to the GOES spacecraft. Methods such as ASSA, ASAP, and MAG4 produce 60-min forecasts but NOAA probabilities are given for a 24-hours time period reported at 00:00 UT. To obtain prediction time series of equal cadence, we assumed the daily NOAA forecast values as the probability for every hour in that day. Most methods produce rather inaccurate forecasts (or no reports at all) for active regions too close to the solar limbs, where the magnetic structure of active regions can't be properly resolved. For this reason, only probabilities that are generated for active regions which longitudinal position is within -70 deg $<$ longitude $<$ 70 deg are included in the time series. Due to the longitudinal constrain, all time series (probabilities and events) are 240 h (10 days) in length.

For a given active region it is straightforward to identify the probabilities from MAG4 and NOAA from the output text files since both methods use the NOAA active-region number labels system. On the other hand, ASSA and ASAP identify and label the active regions present in the solar disk using a non-sequential label number. For these methods, we use the initial position (latitude and longitude) of the active region to track its movement across the solar disk. For missing values, we use two-points interpolation in order to complete the time series.

Probabilistic forecasts given by the ASSA model are reported for a 12 hour prediction window. In order to incorporate the ASSA probabilities into the ensemble forecast, they must be converted to 24-hours predictions. The ASSA model calculates the probabilities following \cite{Bloomfield2012} and \cite{Gallagher2002}: given the (M or X) flare rate $R$ for each McIntosh class, the flaring probability is calculated as

\begin{displaymath}
P_{12h}=1-\exp(-R\Delta t),
\end{displaymath}
\noindent
where $R$ is the average flare rate per 12-hour intervals (see \cite{assa_manual}, page 13) and $\Delta t=$ 12 hours. Since $R$ has units of (12 h)$^{-1}$, the expression above becomes

\begin{displaymath}
P_{12h}=1-\exp(-R),
\end{displaymath}

The 24-hr probability can be computed as

\begin{displaymath}
P_{24h}=1-\exp(-R\Delta t) = 1-\exp(-(R/12hr)24 hr)=1-\exp(-R')
\end{displaymath}
\noindent

where $R'=2R$ is the average flare rate for that McIntosh class for 24-hours intervals. Combining the expressions for $P_{12hr}$ and $P_{24hr}$ we arrive at expression

\begin{displaymath}
P_{24h}=1-(1-P_{12hr})^{2}.
\end{displaymath}
\noindent

On the other hand, as mentioned on Section 2.1, MAG4 predicts the combined rate $R_{\rm M\&X}$ for M- and X- class flares and the rate $R_{\rm X}$ for X-class flares alone. Using $R_{\rm M\&X}$ and $R_{\rm X}$, we calculate the flare rate for only M-class flares as $R_{\rm M}=R_{\rm M\&X}-R_{\rm X}$.

For the events time series we looked for flares that took place within the prediction window, 24 hours from the forecast time. If a flare did take place during this time window, then a value of 100 (true) is assigned to that time in the time series, otherwise a zero (false) value is assigned. When a flare takes place, an entire 24-hours time segment, before the flare time, takes the true value in $E(t)$ (dotted line in Figure \ref{time_series}).

The sample of active regions selected for this study consists of 13 recent active regions with major flaring activity (M- and X-class flares). We selected active regions observed between 2012 and 2014 (see Table \ref{table1}) satisfying two main criteria: 1) active regions producing flares away from the solar limbs (-70 deg $<$ longitude $<$ 70 deg) and 2)  data from all four contributing forecasting methods were available.

Table 1 provides properties of the active regions contained in the sample used in this study. In particular, initial and most complex Mt Wilson and McIntsoh classifications are given as a measure of the active region magnetic field evolution. The sample contains active regions that displayed maximum McIntosh class D, E, or F meaning that they are formed by bipolar sunspot groups.  For a representative statistics, our sample should include a balanced number of active regions from north and south hemispheres, active regions with different levels of flaring activity, and a broader group of Mt Wilson and McIntosh classifications. However, due to the data availability and restrictions in the longitudinal position, our data selection is biased toward flaring active regions.
%
%
%
%
%

\begin{acknowledgments}
We would to thank the developing teams for the models employed in this study: U. of Alabama - Huntsville for MAG4, the Korean Space Weather Center for ASSA, and U. of Bradford - UK for ASAP. Probabilistic data for each model was obtained from their forecast reports and can be accessed through iSWA -- MAG4: \url{http://iswa.ccmc.gsfc.nasa.gov/iswa_data_tree/model/solar/mag4/HMI_NRT-DATA/}, ASSA: \url{http://iswa.ccmc.gsfc.nasa.gov/iswa_data_tree/model/solar/assa/spot/}, and ASAP: \url{http://iswa.ccmc.gsfc.nasa.gov/iswa_data_tree/model/solar/asap/flare-monitor-data/}. We also thank Chris Balch and Rob Steenburgh from SWPC NOAA for providing the NOAA historical data and useful discussions. NOAA forecast reports and historical events reports can be found at \url{ftp://ftp.swpc.noaa.gov/pub/forecasts/daypre/} and \url{http://iswa.ccmc.gsfc.nasa.gov/iswa_data_tree/index/solar/noaa-swpc/events/}, respectively. We also thank D. Falconer, S. Hong, S. Lee, and R. Qahwaji for providing feedback on this investigation.
\end{acknowledgments}

\end{article}
%
%
%
%
%
%

%
%


\begin{table}[htdp]
\begin{center}
\caption{Details on the sample of active region used in this study.}
\begin{tabular}{|c|c|c|c|c|c|c|c|}
\hline
 NOAA AR & Initial date - & Initial Position & X-Class & M-Class & Mt Wilson & McIntosh \\
    \# & Time & [Degrees] & flares \# & flares \# & Class & Class \\ 
\hline\hline
11429 & 2012 Mar. 04 - 00:00 &  18.0/-68.0 & 2 & 11 & $\beta\gamma/\beta\gamma\delta$ & Dkc/Ekc \\
11882 & 2013 Oct. 25 - 12:00 & -08.0/-59.0 & 1 & 4 & $\beta/\beta\gamma\delta$ & Dso/Dkc \\
11890 & 2013 Nov. 04 - 00:00 & -10.0/-62.0 & 3 & 4 & $\beta\gamma/\beta\gamma\delta$ & Ehc/Ekc \\
11944 & 2014 Jan. 03 - 00:00 & -08.0/-64.0 & 1 & 4 & $\beta\gamma/\beta\gamma\delta$ & Fkc/Fkc \\
11967 & 2014 Jan. 29 - 00:00 & -12.0/-67.0 & 0 & 15 & $\beta\gamma/\beta\gamma\delta$ & Eki/Fkc \\
11974 & 2014 Feb. 07 - 00:00 & -12.0/-62.0 & 0 & 10 & $\beta/\beta\gamma\delta$ & Dso/Fkc \\
12002 & 2014 Mar. 09 - 00:00 & -19.0/-64.0 & 0 & 6 & $\beta/\beta\gamma\delta$ & Cao/Ekc \\
12010 & 2014 Mar. 18 - 00:00 & -14.0/-64.0 & 0 & 1 & $\beta/\beta\gamma\delta$ & Bxo/Dac \\
12017 & 2014 Mar. 23 - 00:00 &  09.0/-58.0 & 1 & 2 & $\beta/\beta\gamma\delta$ & Cao/Dac \\
12035 & 2014 Apr. 13 - 00:00 & -16.0/-62.0 & 0 & 1 & $\beta\gamma/\beta\gamma\delta$ & Eai/Ekc\\
12077 & 2014 May  31 - 12:00 & -05.0/-66.0 & 0 & 1 & $\beta/\beta\gamma$ & Cao/Dai \\
12080 & 2014 Jun. 04 - 00:00 & -13.0/-57.0 & 0 & 2 & $\beta/\beta\gamma\delta$ & Bxo/Ekc \\
12087 & 2014 Jun. 12 - 12:00 & -18.0/-56.0 & 0 & 3 & $\beta\gamma/\beta\gamma\delta$ & Dcs/Eac \\
\hline
\end{tabular}
\tablenotetext{a}{The two values of McIntosh and Mt Wilson classes correspond to the initial (Date) state and the maximum display during the evolution. Maximum McIntosh class is determined by the observed flaring probability of the class, see \cite{Bloomfield2012} Table 2.}
\label{table1}
\end{center}
\end{table}


\begin{table}[htdp]
\begin{center}
\caption{Two-by-two contingency table and the performance metrics derived from this table.}
\begin{tabular}{|c|c|c|}
\hline
 & Event & Observed  \\
 \hline
 Event Forecast & Yes & No \\
 \hline
 Yes & a & b \\
 \hline
 No & c & d \\
 \hline
  & & \\
  \hline
  \end{tabular}
  \begin{tabular}{ll}
 PC = (a + d)/n & FAR = b/(a + b)  \\
 PDO = a/(a + c)& n = a + b + c + d \\
 HSS = (a + d - e)/(n - e) & e = [(a+b)(a+c)+(b+d)(c+d)]/n \\
 TSS = (ad - bc)/[(a+c)(b+d)] & \\
\end{tabular}
\label{table2}
\end{center}
\end{table}


\begin{table}[htdp]
\begin{center}
\caption{Table of the attributes and skill for the probabilistic ensemble forecasting method and the ensemble members.}

M-class (Total events\footnotemark: 888, Climatology=0.285, $\overline{QR}=0.204$) \\
\begin{tabular}{|c|c|c|c|c|c|c|}
\hline
 Forecasting method & $\langle P_{i} \rangle$ & $QR$ & $REL$ & $RES$ & $SS$ \\
 \hline
MAG4 & 0.286 & 0.219 & 0.0487 & 0.0329 & -0.0756 \\
ASSA & 0.261 & 0.194 & 0.0311 & 0.0397 & 0.0464 \\
ASAP & 0.197 & 0.212 & 0.0272 & 0.0193 & -0.0398 \\
NOAA & 0.295 & 0.186 & 0.0145 & 0.0363 & 0.0882 \\
Ensemble I & 0.318 & 0.182 & 0.0092 & 0.0307 & 0.1066 \\
Ensemble II & 0.319 & 0.184 & 0.0109 & 0.0299 & 0.1127 \\
Ensemble Equal & 0.321& 0.181 & 0.0076 & 0.0303 & 0.1103 \\
 \hline
\end{tabular}

 X-class (Total events: 169, Climatology=0.054, $\overline{QR}=0.051$)\\
 
 \begin{tabular}{|c|c|c|c|c|c|c|}
 \hline
  Forecasting method & $\langle P_{i} \rangle$ & $QR$ & $REL$ & $RES$ & $SS$ \\
 \hline
MAG4 & 0.159 & 0.061 & 0.0124 & 0.0028 & -0.1870 \\
ASSA & 0.012 & 0.052 & 0.0033 & 0.0025 & 0.0062 \\
ASAP & 0.209 & 0.100 & 0.0642 & 0.0105 & -0.9536 \\
NOAA & 0.058 & 0.056 & 0.0141 & 0.0094 & -0.0966 \\
Ensemble I & 0.099 & 0.054 & 0.0087 & 0.0063 & -0.0472 \\
Ensemble II & 0.079 & 0.052 & 0.0060 & 0.0050 & -0.0647 \\
Ensemble Equal & 0.100 & 0.054 & 0.0079 & 0.0050 & -0.0602 \\
 \hline
 \end{tabular}
\tablenotetext{a}{The total number of forecasts in our study is 3120. $QR$, $REL$, and $RES$ stand for accuracy, reliability, and resolution, as defined in the text, and $SS$ is the skill of the ensemble compared to the climatology. $\langle P_{i} \rangle$ corresponds to the average probability value for each method. $\overline{QR}$ is the accuracy of the climatological predictor.}
\tablenotetext{1}{One event is defined as a (M/X) flare observed within the 24-hours prediction window from that hourly forecast (see Appendix A). For the number of flares observed in our sample, see Table \ref{table1}.}
\label{table3}
\end{center}
\end{table}
\end{document}